\documentclass[%
preprint,
 amsmath,amssymb,
 aps,
prf,
]{revtex4-2}

\usepackage{graphicx}
\usepackage{dcolumn}
\usepackage{bm}
\usepackage{comment}
\usepackage{hyperref}
\hypersetup{hidelinks}
\usepackage{subcaption}
\usepackage[dvipsnames]{xcolor}
\usepackage{booktabs}
\usepackage{adjustbox}
\usepackage{tabulary}

\begin{document}

\preprint{APS/123-QED}

\title{Self-organization in a stably stratified, valley-shaped enclosure heated from below}

\author{Patrick J. Stofanak}
\author{Cheng-Nian Xiao}
\author{Inanc Senocak}%
 \email{senocak@pitt.edu}
\affiliation{%
 Department of Mechanical Engineering and Materials Science,\\
 University of Pittsburgh, Pittsburgh, PA 15261, USA
}%


\date{\today}

\begin{abstract}
We observe the spontaneous onset of three-dimensional motion from a quiescent, purely conductive state of a stably stratified fluid in a V-shaped enclosure heated from below, which ultimately self-organizes into a two-dimensional steady state without any external forcing to the initial configuration. 
We identify a dominant three-dimensional instability through modal stability analysis.  Direct numerical simulations confirm this instability but also reveal that, after an initial period of spontaneous three-dimensional growth, the flow gradually self-organizes into a steady two-dimensional state without external intervention. This self-organization manifests consistently for any arbitrary infinitesimal three-dimensional disturbance to the initial quiescent configuration. We demonstrate that the mechanism driving this self-organization is the increasing dominance of viscous dissipation over buoyant production of disturbance kinetic energy at later stages of flow evolution from the initial quiescent state. 
Our investigation reveals a flow scenario in which the most natural transition pathway to the final state involves passing through an intermediate state with a higher dimension than the final state itself. 
Specifically, our final flow state is less complex than the three-dimensional most unstable eigenvector predicted by linear stability analysis. 
\textcolor{black}{We demonstrate that the entire flow evolution remains non-turbulent throughout and closely aligns with results from linear stability analysis, distinguishing the present flow dynamics from transient chaos, which also features complex transient states that eventually converge to a less complex final state.}
\end{abstract}

\maketitle


\section{\label{sec:level1} Introduction} 


Self-organization refers to the spontaneous emergence of spatial, temporal, or spatiotemporal patterns in complex, dissipative systems driven away from thermodynamic equilibrium. Such behaviors are of fundamental importance in a variety of fields, as they exemplify how order can arise in seemingly chaotic environments. In fluid systems, self-organization phenomena encompass a broad spectrum of scales and applications, particularly in geophysical and astrophysical contexts. Notable examples include large-scale structures such as the formation of galaxies and Jupiter’s red spot \citep{aschwanden2018order}, as well as pattern formation like hexagonal convection cells in Rayleigh-B\'{e}nard convection \citep{bodenschatz2000recent} and the formation of laminar-turbulent bands in Taylor-Couette flows \citep{tuckerman2020patterns}. 
These phenomena highlight the inherent interplay between pattern formation and the transport of mass, momentum, and energy within the system. The feedback mechanisms governing these processes are often nonlinear and pose significant challenges to understanding and predicting their behavior.

In this study, we investigate the initiation of convection in a stably stratified, quiescent fluid contained within a V-shaped triangular cavity heated from below. We employ linear stability analysis (LSA) and direct numerical simulations (DNS) to explore this phenomenon. Our findings reveal that this seemingly simple convective system exhibits \textit{spontaneous} three-dimensional (3D) growth from an initially quiescent state, which then self-organizes into a steady two-dimensional (2D) state without any external stimulation to the initial configuration.

In dynamical systems, the system's disorder is typically based on the spatial and temporal dimensionality of the solution.
Traditionally, if an initially ordered state is dynamically unstable, it is expected to become progressively more disordered when perturbed \citep{gollub1995order}. This type of evolution is common in fluid systems, such as transition to turbulence through a sequence of instabilities \citep{schmid2002stability}. 
However, our findings contradict this expectation, as the base state initially exhibits instability to a 3D mode, as predicted by LSA and further confirmed by DNS. This intermediate 3D flow then self-organizes into a final steady 2D state on its own. This constitutes a unique path to self-organization, deeply tied to the system's dynamical stability. 

In linear stability theory, a state is said to be unstable when there is an unstable eigenvalue to the linearized system. For some flow configurations, the prediction of the linear theory matches well with experiments, for example in Taylor-Couette or Rayleigh-B\'{e}nard flows. However, when the linearized operator is highly non-normal, the predictions of the linear theory can fail to match experiments, and significant transient amplification of initial disturbances are possible even when all eigenvalues are stable \citep{trefethen1993hydrodynamic}. 
Such transient growth behavior has been widely studied in shear-dominated flows such as plane Couette and plane Poiseuille flows, in which small, finite-size, initial 3D disturbances undergo growth of several orders of magnitude before either exponentially decay back to the initial state or transition to turbulence \citep{reddy1993energy}, an area known as non-modal analysis \citep{schmid2007nonmodal}. However, in our problem, non-modal behavior is not an issue due to the strong parallel between our set-up and that of Rayleigh-B\'{e}nard convection which has been proven to be modal up to the first bifurcation. We also provide evidence from direct numerical simulations (DNS) that support our assumption of modality. 

Self-organization is not uncommon in fluid systems. 
However, self-organization manifests uniquely with differing causes underlying the flow system. Here, we review a number of notable examples of self-organization in fluid systems, which bear some resemblance to the phenomenon observed in our case but with important differences in problem settings and external forcing.
One example is the case of 2D decaying turbulence in bounded domains, which exhibit an initial `spin-up' of angular momentum before the flow self organizes into long-lived vortices that decay over time \citep{clercx1998spontaneous,maassen2002self}. 
Other studies observe relaminarization of turbulence due to an external change imposed on the disordered state. \citet{xu2023wall} study the case of Rayleigh-B\'{e}nard convection with imposed wall shear and find that an initially turbulent state can be relaminarized with an increase in wall shear, and that the subsequent laminar state achieves more efficient heat transfer.
\citet{kuhnen2018destabilizing} observe relaminarization of turbulent pipe flow in experiments and simulations when the flow profile is flattened by adding perturbations in experiments and adding a forcing term in the simulations. A similar effect was observed by heating of the pipe walls as well \citep{marensi2021suppression}. 

Pattern formation of laminar-turbulent bands as well as complete laminarization are known to occur in a variety of wall-bounded shear flows, such as plane Couette and Taylor-Couette flow \citep{tuckerman2020patterns}, but crucially this involves lowering the Reynolds number from a previously turbulent state \cite{barkley2005computational}.
\citet{kaiser2020stages} describe the decay of solid-body rotation flow within an impulsively stopped rotating cylinder, which includes the transition from an initially laminar boundary layer flow to turbulence, before decaying back to laminar flow again. 
However, in each of these cases, the self-organization behavior can be explained by an initial large perturbation of the flow or an added external forcing.

In 2D decaying turbulence, because there is nothing driving the turbulence, the transition from an initially complex state to a more ordered state at later times is expected. In the examples of relaminarization in pipe flow, an external change is imposed on the initial turbulent, disordered state, and this change has the effect of stabilizing the flow and causing the collapse of turbulence to a more ordered laminar state. Finally, while the case of the impulsively stopped cylinder may be said to follow a similar transition scenario to the results described here (namely the transition from an ordered state to a more disordered state before returning to an ordered state again), we stress that the stopping of the cylinder rotation represents an initial finite amplitude disruption to the flow field, and the flow field at this initial time is not in equilibrium, as it is initially in our example. 

Additionally, a number of studies have investigated the two-dimensionalization of 3D turbulence due to the difference in energy cascade that is observed between 3D and 2D turbulence \citep{ecke20172d}. Various mechanisms have been used to show this two-dimensionalization, including the use of thin, quasi-2D geometries, global rotation of the fluid, or the presence of a strong magnetic field. For example, \citet{benavides2017critical} simulate thin layer turbulence and shows that a critical value exists at which the energy cascade transitions to a purely inverse cascade, indicative of 2D turbulence. For the case of magnetohydrodynamics, \citet{gallet2015exact} consider a flow driven by a purely horizontal body force and subject to a vertical magnetic field. They prove that for certain parameter values, the flow becomes 2D in the long time limit regardless of the initial conditions, and for another set of parameter values, the flow remains 2D under infinitesimal perturbation in the third direction. The existence of the same `absolute 2D' and `linearly 2D' states was also proven for the case of strong global rotation \citep{gallet2015exact1}. Additionally, the transition from 3D to 2D turbulence has been shown for Taylor-Couette flow under the presence of a strong magnetic field \citep{zhao2014transition}.

\textcolor{black}{
The co-existence of a complex transient state with a simple asymptotic attractor is not uncommon in nonlinear dynamical systems, and it is often studied in instances of transient chaos \citep{lai2011transient}. Transient chaos is most often manifested by spatio-temporal chaotic behavior, exhibiting sensitivity to initial conditions, which eventually collapses to a temporally periodic or time-independent attractor \citep{tel2008chaotic,lai2011transient}. 
This behavior has been shown to be caused by the existence of a chaotic saddle, or a non-attracting chaotic set in the phase space in both low-dimensional and high-dimensional spatially extended systems \citep{nusse1989procedure,rempel2007origin}. Transient chaos has been observed in a number of fluid systems. In Rayleigh-B\'{e}nard (RB) convection, \citet{chimanski2016off} observe intermittent periods of chaotic behavior and temporally periodic behavior in 3D RB simulations. \citet{oliveira2024transition} observe similar intermittency in the presence of rotation and a magnetic field.}

\textcolor{black}{Transient chaos has also been used to comprehend the transition to turbulence in shear flows. For example, the transition to turbulence in pipe flow has been shown to be caused by a chaotic saddle \citep{faisst2004sensitive,eckhardt2007turbulence}. This leads to states that appear chaotic or turbulent for some time before collapsing back to the laminar state (or relaminarizing) \citep{avila2023transition}. This is in contrast to the prior examples of relaminarization, in which a perturbation is made to the flow \citep{xu2023wall,kuhnen2018destabilizing}. Instead the relaminarization of turbulence in pipe flow is due to the non-attracting nature of the chaotic saddle, and the attracting laminar state. Similar relaminarization of transient turbulence due to a chaotic saddle has also been observed for plane Couette flow \citep{kreilos2012periodic,lustro2019onset}, and plane Poiseuille flow \citep{zammert2015crisis}.  }


In the following, we describe flow self-organization in a valley enclosure with a stably stratified fluid heated from below, starting from the initial 3D growth of a modal disturbance through the eventual decay of three-dimensionality to the final 2D steady state. 
We explain this unexpected evolution through analysis of the disturbance kinetic energy budget terms as well as through the secondary stability of the 2D steady state. 

\section{Problem description} 

\begin{figure}
\centering
\includegraphics[width=0.55\textwidth]{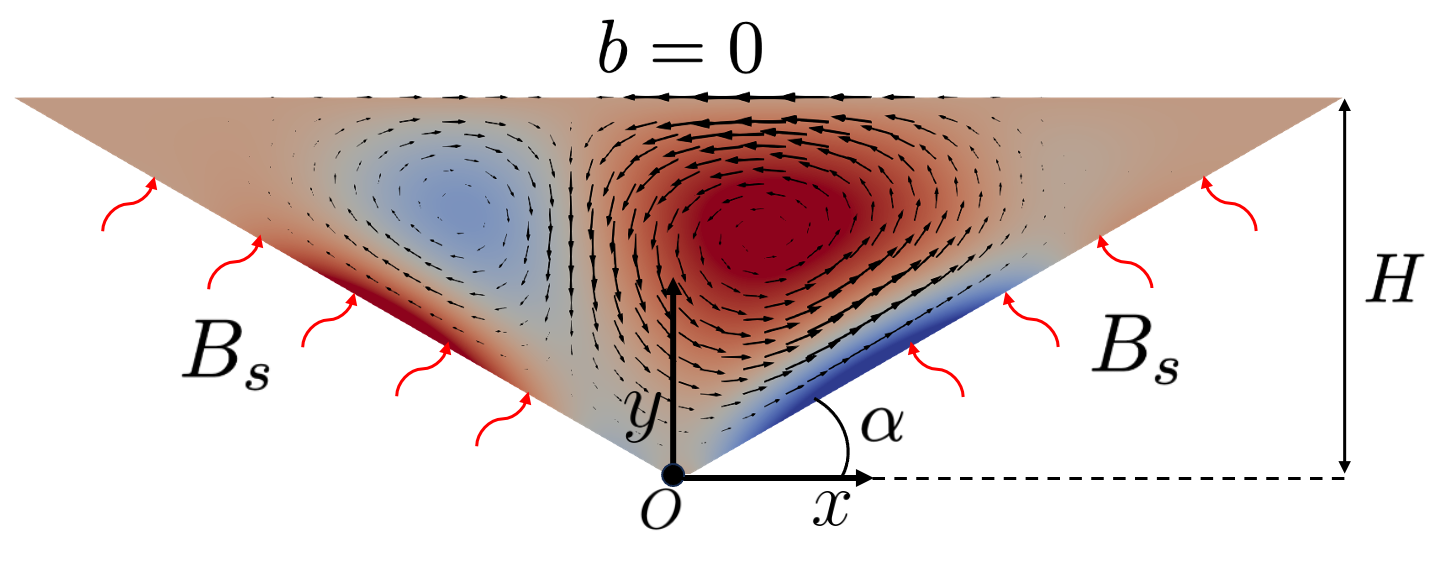}
\caption{\label{fig:schematic} Schematic of the computational domain, buoyancy boundary conditions, and key parameters, illustrated with vorticity and velocity vectors of the 2D asymmetric steady state. The $x$ and $y$-axes are shown at the origin $O=(0,0,0)$ with the positive $z$-axis being out of the page. All simulations and visualisations adopt these axes.}
\end{figure}

We examine a simple model of stably stratified flow within a V-shaped cavity heated from below, as illustrated in Figure \ref{fig:schematic}. The figure details key parameters and boundary conditions as well. The V-shaped cross-section is oriented in the $x-y$ plane, featuring horizontal $u$ and vertical $v$ velocity components, while the $w$ velocity component lies in the homogeneous $z$ direction. The walls of the triangular enclosure are inclined at a slope angle $\alpha$ of $30^{\circ}$. Buoyancy is defined by the equation $b = (g/\Theta_r) (\Theta - \Theta_e(y))$, where $\Theta$ represents the potential temperature, $\Theta_r$ denotes a reference potential temperature, and $\Theta_e$  signifies the potential temperature of the ambient environment. We impose a constant stable stratification, characterized by the buoyancy frequency $N = \sqrt{(g/\Theta_r) \partial \Theta_e/\partial y}$. Consequently, $b$ represents a perturbation from the linear background stratification defined by the constant $N$.

The governing equations are represented by the Navier-Stokes equations, formulated under the Oberbeck-Boussinesq approximation:
\begin{equation} \label{eq:NS1}
    \nabla \cdot \mathbf{u} = 0,
\end{equation}
\begin{equation} \label{eq:NS2}
    \frac{\partial \mathbf{u}}{\partial t} + \mathbf{u} \cdot \nabla \mathbf{u} = - \nabla p + \nu \nabla^2 \mathbf{u} + b \mathbf{g},
\end{equation}
\begin{equation} \label{eq:NS3}
    \frac{\partial b}{\partial t} + \mathbf{u} \cdot \nabla b = \beta \nabla^2 b - N^2 \mathbf{g} \cdot \mathbf{u},
\end{equation}
where $\nu$ is the kinematic viscosity, $\beta$ is the thermal diffusivity, and $\mathbf{g}$ is the effective gravity vector $\mathbf{g} = [0, 1, 0]$. 
The cavity is heated from the two sloping bottom walls with a constant buoyancy flux $B_s = \beta \partial b / \partial n$ and no-slip velocity boundary conditions, and a constant $b = 0$ is imposed on the flat top boundary with free-slip velocity conditions. \textcolor{black}{Boundary conditions in the $z$ direction are periodic.} Given these conditions, a quiescent, zero-flow state exists with exact solutions for normalized buoyancy and pressure profiles given by:
\begin{equation} \label{eq:base_flow}
p(y) = -\frac{1}{2} \left( y - 1 \right)^2, \qquad b(y) = 1 - y .
\end{equation}
The exact solutions for pressure and buoyancy are normalized by the following scales, along with the length, velocity, and time scales: 
\begin{equation}
    l_0 = H, \quad u_0 = \sqrt{\frac{B_s}{N}}, \quad b_0 = \frac{B_s H}{\beta \cos \alpha}, \quad p_0 = \frac{B_s H^2}{\beta \cos \alpha} , \quad t_0 = \frac{H^2}{\nu},
\end{equation} 
where $H$ is the height of the valley geometry. Hereafter, all variables are nondimensionalized by the appropriate scale.

The flow is controlled by the following four dimensionless parameters: 
\begin{equation}
    \Pi_s = \frac{B_s}{\beta N^2}, \quad \Pi_h = \frac{N H^2}{\beta}, \quad Pr = \frac{\nu}{\beta}, \quad \alpha,
\end{equation}
where $\Pi_s$ is the stratification perturbation parameter, a measure of the strength of surface buoyancy flux relative to the background stratification, $\Pi_h$ is the buoyancy number, a ratio of the thermal diffusion and stratification time scales, along with Prandtl number $Pr$, and slope angle $\alpha$. In our results, we only vary $\Pi_s$ and $\Pi_h$, setting the Prandtl number to 7 and the slope angle to $30^{\circ}$. \textcolor{black}{These values were chosen to parallel the prior experimental work done by \citet{princevac2008morning}, as well as by the fact that this set-up is amenable to future experimental investigation.}

\subsection{Linear stability analysis}
We linearize the Navier-Stokes equations around an arbitrary 2D base flow given by $(U(x,y), V(x,y), P(x,y), B(x,y))$ and assume disturbances $\mathbf{q}$ of the form
\begin{equation} \label{eq:disturbance}
    \mathbf{q}(x, y, z, t) = \mathbf{\hat{q}} \exp \left( ik_z z + \omega t \right),
\end{equation}
where $\mathbf{\hat{q}} = [\hat{u}(x,y), \hat{v}(x,y), \hat{w}(x,y), \hat{p}(x,y), \hat{b}(x,y)]$ represents the vector of 2D disturbance quantities which vary only in the $x$-$y$ plane,  $k_z$ represents the wavenumber in the homogeneous $z$-direction, and $\omega$ represents the temporal growth rate. 
Substituting the disturbance of Eq. \ref{eq:disturbance} into the linearized Navier-Stokes equations yields 
\begin{align}
\frac{\partial \hat{u}}{\partial x} + \frac{\partial \hat{v}}{\partial y} + i k_z \hat{w} = 0,\label{eq:lsa1}\\
\omega \hat{u} + U \frac{\partial \hat{u}}{\partial x} + V \frac{\partial \hat{u}}{\partial y} + \hat{u} \frac{\partial U}{\partial x} & 
+ \hat{v} \frac{\partial U}{\partial y} =  - \frac{\partial \hat{p}}{\partial x} + \nu \left(   \frac{\partial^2\hat{u}}{\partial x^2}+\frac{\partial^2\hat{u}}{\partial y^2} - k_{z}^{2} \hat{u} \right),\\ 
\omega \hat{v} + U \frac{\partial \hat{v}}{\partial x} + V \frac{\partial \hat{v}}{\partial y} + \hat{u} \frac{\partial V}{\partial x} & 
+ \hat{v} \frac{\partial V}{\partial y} = -\frac{\partial \hat{p}}{\partial y} + \nu \left( \frac{\partial^2\hat{v}}{\partial x^2}+\frac{\partial^2\hat{v}}{\partial y^2} - k_{z}^{2} \hat{v} \right) + \hat{b},\\
\omega \hat{w} + U \frac{\partial \hat{w}}{\partial x} + V \frac{\partial \hat{w}}{\partial y} &
= -ik_z \hat{p} + \nu \left( \frac{\partial^2\hat{w}}{\partial x^2}+\frac{\partial^2\hat{w}}{\partial y^2} - k_{z}^{2} \hat{w} \right) ,\\
\omega \hat{b} + U \frac{\partial \hat{b}}{\partial x} + V \frac{\partial \hat{b}}{\partial y} + \hat{u} \frac{\partial B}{\partial x} &
+ \hat{v} \frac{\partial B}{\partial y}
= \beta \left( \frac{\partial^2\hat{b}}{\partial x^2}+\frac{\partial^2\hat{b}}{\partial y^2} - k_{z}^{2} \hat{b} \right) - N^2 \hat{v} \label{eq:lsa_last}
\end{align} 
The above equations can be written as a generalized eigenvalue problem of the form:
\begin{equation} \label{eq:eig_prob}
    \mathbf{A}(k_z) \mathbf{\hat{q}} (x, y) = \omega \mathbf{B}(k_z) \mathbf{\hat{q}} (x, y).
\end{equation}
Solving the above eigenvalue problem represents a linear stability analysis for a given 2D base flow in the valley geometry and a given homogeneous spanwise extent defined by $k_z$.
Given the definition of our disturbance, the real part of the eigenvalue, $\mathrm{Re}(\omega)$, represents the growth rate of the disturbance, with a positive value indicating exponential growth and a negative value indicating exponential decay. The imaginary part of the eigenvalue, $\mathrm{Im}(\omega)$, represents the frequency of the associated disturbance, with $\mathrm{Im}(\omega)=0$ representing a stationary mode. For each eigenvalue, the associated eigenvector, $\mathbf{\hat{q}} = [\hat{u}, \hat{v}, \hat{w}, \hat{p}, \hat{b}]^{T}$ represent the shape of the disturbance corresponding the growth rate and frequency. 


All simulations are carried out with the spectral/hp element code N\textsc{ektar}++ \citep{cantwell2015nektar++,moxey2020nektar++}. 
Our simulations employ a discretization of 31 elements along each bottom wall and a polynomial order of 4 within each element, and the spanwise direction incorporates a Fourier expansion with 16 modes per nondimensional length scale. For the LSA, the eigenvalue problem is solved numerically using the modified Arnoldi method implemented in Nektar++.
The Krylov subspace dimension was varied between 32 and 512, and all simulations were run until several leading eigenvalues converged to at least a residual of $10^{-6}$. We perform additional simulations increasing the number of eigenvalues to converge, but no other unstable eigenmodes were found besides those described here. The eigenvalues and associated residuals from the Arnoldi method for one characteristic case are provided in Table \ref{tab:table1}.
We validated the results of our Nektar++ simulations, including both LSA and DNS, through several ways, including the Nektar++ tutorials which offer a variety of benchmark flow problems, as well as comparison to known results for Prandtl slope flow \citep{xiao2019stability,xiao2022speaker}. Specifically, we run LSA for the 1-D Prandtl analytical base flow and our results closely match those presented in \citet{xiao2019stability} which utilized a separate in-house code to compute the eigenvalues. Additionally, for our valley case, we test the sensitivity of our LSA to mesh refinement by increasing the p order from 4 to 8 and the growth rate of our leading eigenvalue changes by less than 0.1\%. For these reasons, we can have confidence in our numerical results.

\begin{table}
\centering
\begin{tabular}{l|ccc}
              & Growth rate, $\mathrm{Re(\omega)}$ & Frequency, $\mathrm{Im(\omega)}$ & Residual \\ \hline
3D Symmetric  & $6.5720 \times 10^{-2}$ & 0         & $1.3430 \times 10^{-13}$ \\
3D Asymmetric & $5.1492 \times 10^{-2}$ & 0         & $7.4241 \times 10^{-8}$ \\
2D Asymmetric & $5.4919 \times 10^{-2}$ & 0         & $9.1542 \times 10^{-8}$ \\
2D Symmetric  & $4.7026 \times 10^{-2}$ & 0         & $9.7537 \times 10^{-7}$ 
\end{tabular}
\caption{\label{tab:table1} Growth rate, frequency, and residual for each unstable eigenvalue from LSA with $\Pi_s = 0.9$, $\Pi_h = 1500$, and $L_z = 1.6$.}
\end{table}

\section{Results}

\subsection{Linear stability analysis}

\subsubsection{Primary linear stability analysis}

\textcolor{black}{We consider the case with parameter values $\Pi_s = 0.9$, $\Pi_h = 1500$, and $L_z = 1.6$. This set of parameter values was chosen because it is relatively close to the critical value for instability of $\Pi_s = 0.871$, $\Pi_h = 1500$, As a result, it exhibits a slow and orderly transition compared to states with larger parameter values, making it more amenable to detailed analysis.} \textcolor{black}{However, we note that while this study focuses on a specific set of parameters, we observed similar behavior for $ 0.872 \lessapprox \Pi_s \lessapprox 1.3$ at fixed $\Pi_h = 1500$, and for $ 1000 \lessapprox \Pi_h \lessapprox 4000$ at a fixed $\Pi_s = 0.9$.}

\begin{figure}
\centering
\includegraphics[width=0.75\textwidth]{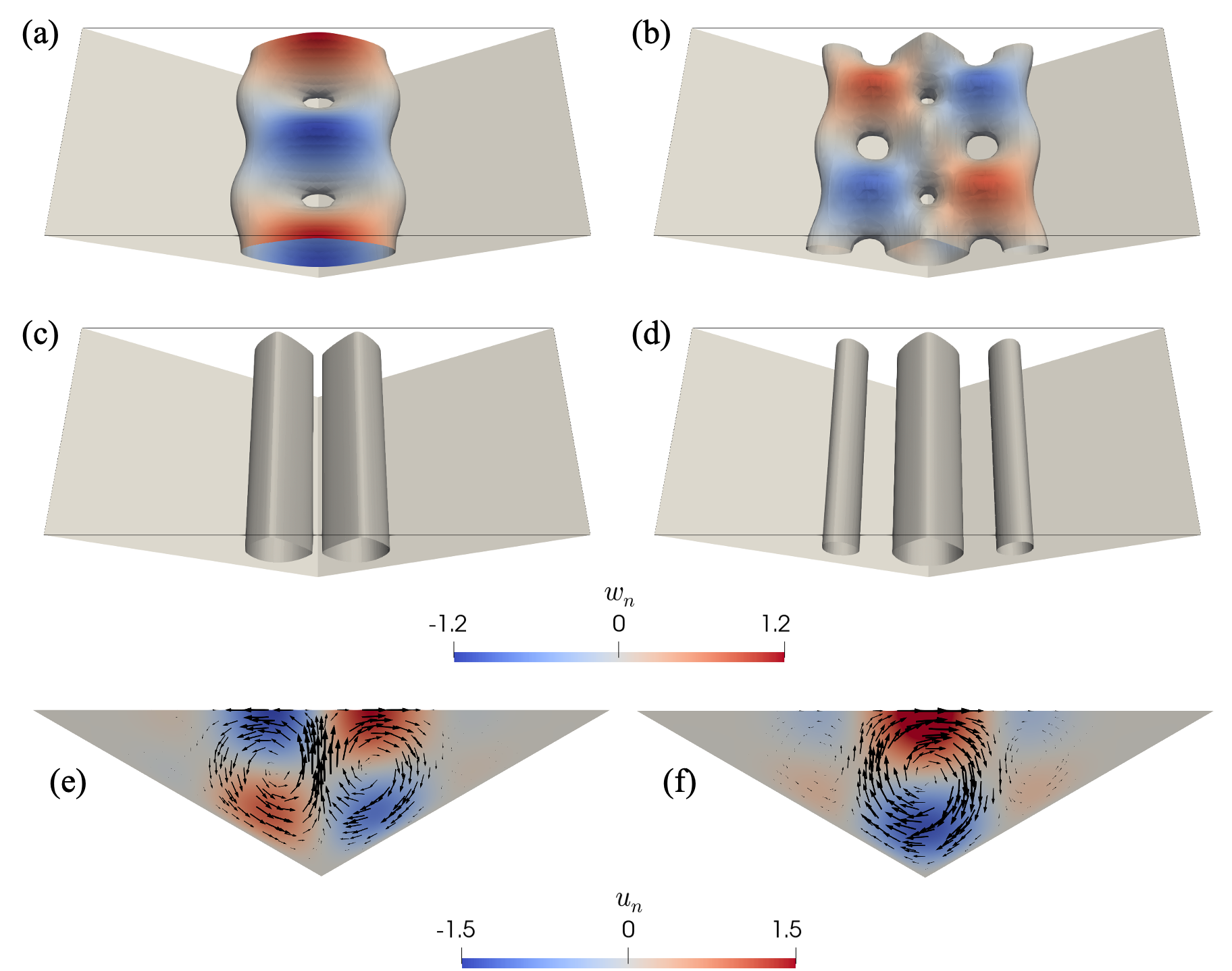}
\caption{\label{fig:eigenvector} Visualization of the four unstable eigenvectors found from LSA: (a) 3D symmetric eigenmode, (b) 3D asymmetric mode, (c) 2D symmetric mode in 3D geometry, (d) 2D asymmetric mode in 3D geometry, (e) 2D symmetric mode 2D velocity profile, and (f) 2D symmetric mode 2D velocity profile. (a-d) show Q-criterion set to approximately 4\% of the maximum value, and are colored by spanwise velocity $w$. (e-f) show 2D velocity vectors and colored by $u$ velocity. Eigenvectors shown are for $\Pi_s = 0.9$, $\Pi_h = 1500$, and $L_z = 1.6$, and correspond to the eigenvalues shown in the spectrum of Figure \ref{fig:lsa_results}(a) and Table \ref{tab:table1}. }
\end{figure}

We first perform linear stability analysis with the quiescent, pure-conduction base flow defined in Eq. \ref{eq:base_flow}, and we find two unstable 3D eigenmodes. The eigenvectors 
are visualized in Figure \ref{fig:eigenvector}(a) and (b) with the contours of the Q-criterion colored by the spanwise velocity $w$. The first mode consists of two circulation rolls intertwined along the $z$-direction, and the second mode consists of three circulations intertwined along the $z$-direction. We note that the first eigenmode in Figure \ref{fig:eigenvector}(a) satisfies the only symmetry of our valley geometry, namely the reflection symmetry about the vertical $y$-axis which can be defined by the transformation
\begin{equation}
\left[u, v, w, b, p \right] \left(x, y \right) \mapsto \left[-u, v, w, b, p \right] \left(-x, y \right).
\end{equation}
The second mode in Figure \ref{fig:eigenvector}(b) does not satisfy this symmetry. Therefore, we refer to these modes as the 3D symmetric and 3D asymmetric modes respectively. Henceforth the use of `symmetric' and `asymmetric' refers to the reflection symmetry about the vertical axis defined by the transformation above. 

LSA of this problem in purely 2D reveals two primary instabilities from the pure-conduction base state, each of which continue to exist in the current stability analysis with $k_z = 0$. These two 2D eigenmodes are visualized in the 3D valley geometry in Figure \ref{fig:eigenvector}(c) and (d) with the Q-criterion, and their 2D velocity profiles are visualized in Figure \ref{fig:eigenvector}(e) and (f).
The first, which we refer to as the 2D symmetric mode, consists of two equal and opposite circulation rolls on either side of the cavity. The second, the 2D asymmetric mode, consists of three circulation rolls, but one of which is dominant and occupies the center of the cavity. Similar to the 3D modes, the 2D symmetric mode satisfies the reflection symmetry of the valley, whereas the 2D asymmetric mode does not. We note that the 3D and 2D asymmetric eigenmodes can also be said to satisfy a reflection symmetry about the vertical $y$-axis defined by the transformation
\begin{equation}
\left[u, v, w, b, p \right] \left(x, y \right) \mapsto \left[u, -v, -w, -b, -p \right] \left(-x, y \right).
\end{equation}
However, this is only a symmetry of the linearized equations, and thus this symmetry must be broken in nonlinear Navier-Stokes simulations. Therefore these eigenmodes are referred to as asymmetric because their use as an initial condition in nonlinear simulations always results in an asymmetric state. More detailed analysis of the symmetries and resulting steady states of these 2D instabilities can be found in \citet{stofanak2024unusual}. We note that while this prior paper explores the same setup of stably stratified flow in a triangular cavity, here our focus is on the observed self-organization and unique route to the steady state solutions, whereas the focus of this prior paper was on the unique bifurcation diagram when only the 2D steady states are considered.

\begin{figure}
\centering
\includegraphics[width=0.49\textwidth]{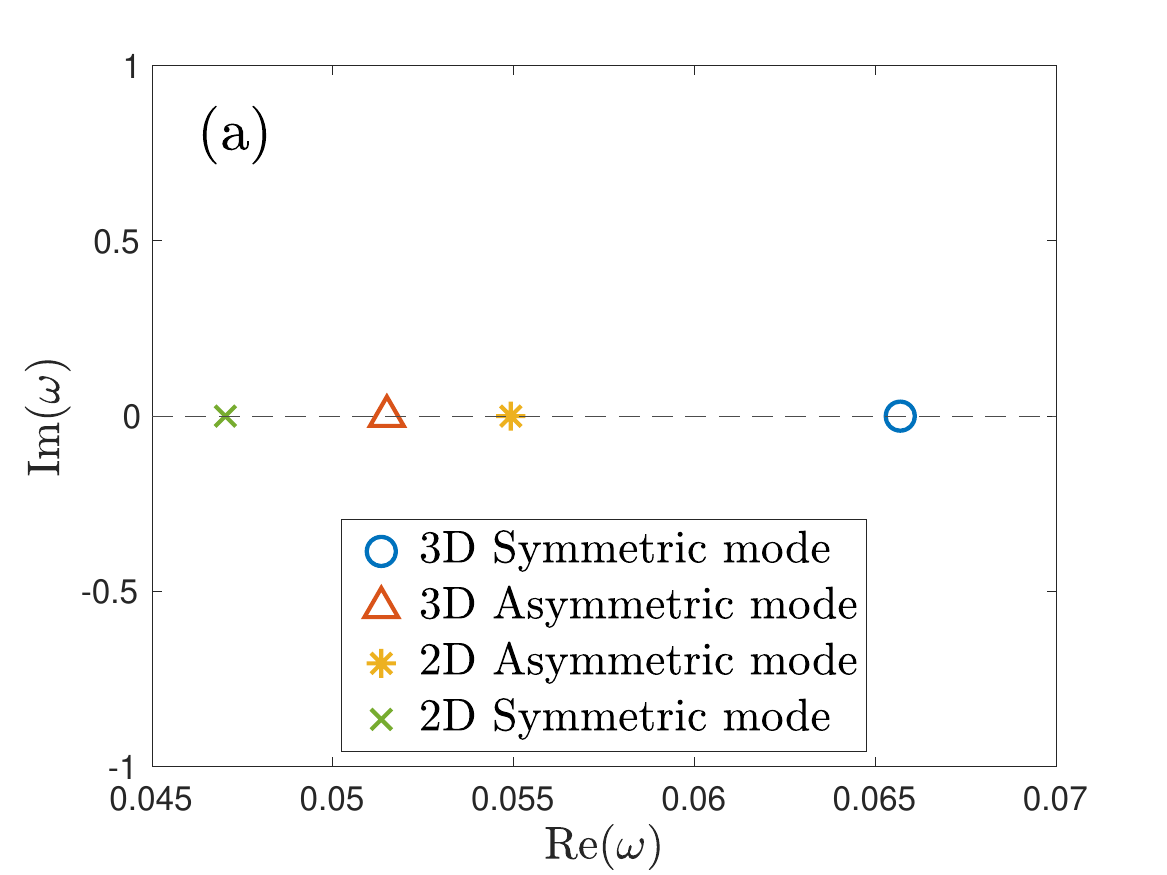}
\includegraphics[width=0.49\textwidth]{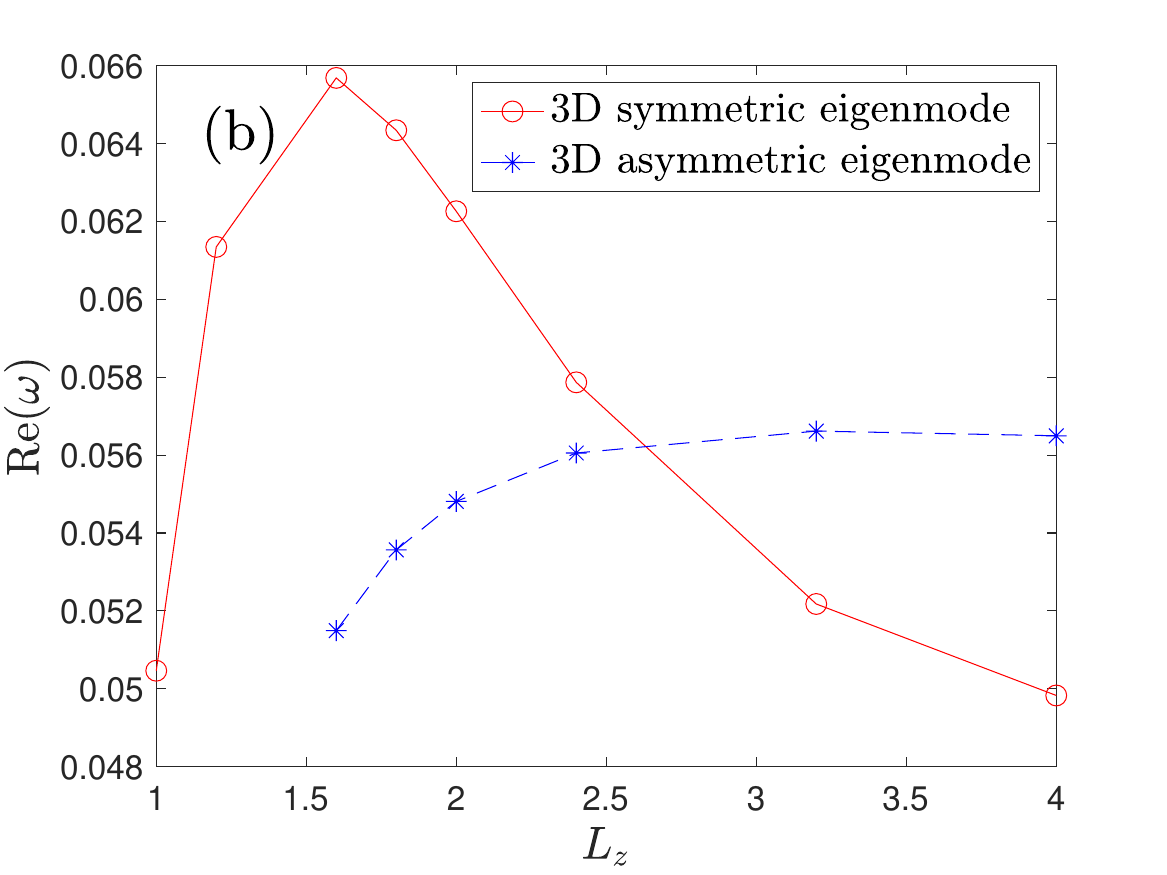}
\includegraphics[width=0.49\textwidth]{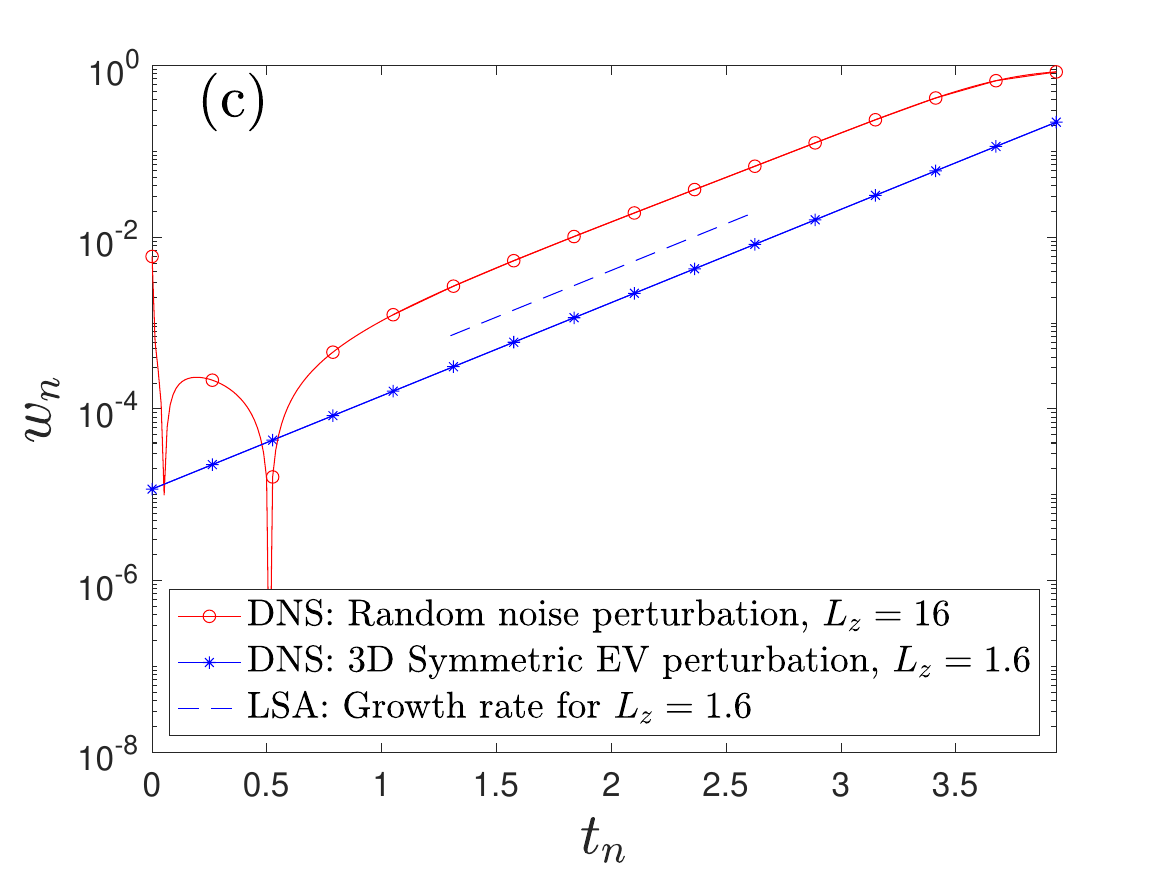}
\caption{\label{fig:lsa_results} (a) Eigenvalue spectrum for $L_z = 1.6$ showing the four unstable eigenvalues. (b) Growth rate of 3D eigenmodes versus $L_z$ with $\Pi_s = 0.9$, $\Pi_h = 1500$, all results from LSA. (c) Comparison of LSA growth rate with two DNS cases: one with initial zero state with random noise perturbation with $L_z = 16$ (red circles), and one with symmetric eigenvector perturbation at $L_z = 1.6$ (blue asterisks), $w$ velocity probed at point (0.0, 0.9, 0) for both cases. The flow fields during the early evolution of the $L_z = 1.6$ and $L_z = 16$ cases are shown in Figure \ref{fig:wSquared}(b) and \ref{fig:Qcrit_16Lz}(b) respectively. } 
\end{figure}

Figure \ref{fig:lsa_results}(a) shows the eigenvalue spectrum for the case of $\Pi_s = 0.9$, $\Pi_h = 1500$, and $L_z = 1.6$, which shows the growth rate and frequency for the four unstable eigenmodes previously described. Our LSA converges to only these four eigenvalues, while any additional eigenvalues that are found to be spurious are discarded. The growth rate $\mathrm{Re(\omega)}$, frequency $\mathrm{Im(\omega)}$, and residual for each eigenvalue is given in Table \ref{tab:table1}. From Figure \ref{fig:lsa_results}(a), it can be seen that the 3D symmetric mode is the dominant mode, with the largest growth rate, and that the temporal frequency of all the modes are zero.
Additionally, the dependence of the growth rate of the two 3D modes on the domain extent in the homogeneous direction $L_z$ is shown in Figure \ref{fig:lsa_results}(b). The optimal wavelength of the 3D symmetric mode is approximately 1.6, and the growth rate of the 3D asymmetric mode exceeds that of the 3D symmetric mode for $L_z \gtrsim 2.6$. However, because the growth rate of the 3D symmetric mode is maximum at small wavelengths and exceeds that of the 3D asymmetric mode at larger wavelengths, we expect to only observe the 3D symmetric mode for any $L_z$ in DNS. Our DNS results support this, and for this reason we focus on the 3D symmetric mode for the remainder of this paper.


We compare the modal LSA results with the initial growth rate obtained from DNS beginning with a disturbance consisting of a small multiple of the 3D symmetric eigenvector with $L_z = 1.6$, as well as small amplitude random noise with a large spanwise domain of $L_z = 16$. Figure \ref{fig:lsa_results}(c) shows that the growth rate predicted by LSA, the dashed line, matches the initial exponential growth of the two DNS with different starting perturbations. This consistent trend supports the modal behavior of the instability due to the fact that even at large $L_z$ and with random initial conditions, the initial growth rate of our DNS matches with the most unstable mode found in our modal stability analysis.



\subsubsection{Two-dimensional steady states and secondary stability analysis} \label{sec:2d}

\begin{figure}
\centering
\includegraphics[width=0.95\textwidth]{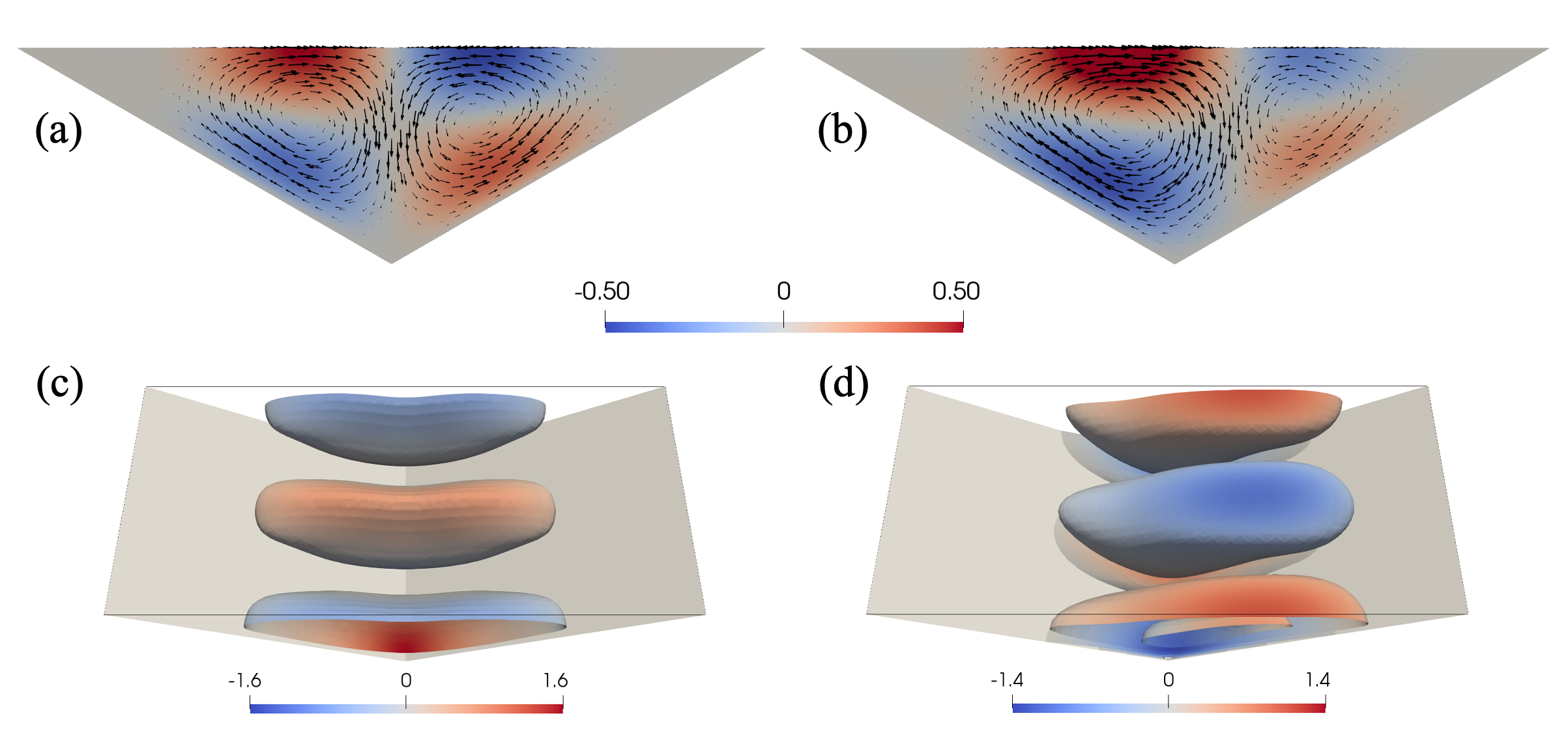}
\caption{\label{fig:2dSteadyStates} Visualization of the (a) 2D symmetric and (b) 2D asymmetric steady states obtained from 2D Navier-Stokes simulations, colored by $u$ velocity for $\Pi_s = 0.9$, $\Pi_h = 1500$, and eigenvectors for the corresponding secondary instabilities to the (c) 2D symmetric state and (d) 2D asymmetric state, both colored by $w$ velocity. The eigenvector in (c) is shown with the Q-criterion contours whereas (d) is shown with $z$-vorticity contours.}
\end{figure}

We first review the steady states arising from the 2D symmetric and asymmetric eigenmodes by performing 2D Navier-Stokes simulations. Each of these eigenmodes leads to a corresponding steady state solution, including a symmetric steady state with two equal and counter-rotating circulation rolls in the center of the valley, and an asymmetric steady state with one stronger roll closer to the center of the valley and smaller, counter-rotating rolls in the corners. The 2D visualization of each of these steady states is depicted in Figure \ref{fig:2dSteadyStates}(a-b). We obtain the 2D steady states through simulations of a purely 2D domain, or through 3D simulations with initialization of only a 2D disturbance. More details of all possible steady-state solutions and the resulting bifurcation diagram for this case is provided in \cite{stofanak2024unusual}.


In addition, we perform 3D secondary stability analysis with each of the 2D symmetric and 2D asymmetric states as the base flow. This is carried out for the case of $\Pi_s = 0.9$, $\Pi_h = 1500$, and $L_z = 1.6$, and the resulting growth rates and frequencies are reported in Table \ref{tab:table2}.
For each of the 2D steady states, we obtain a 3D eigenmode from LSA which are visualized in Figure \ref{fig:2dSteadyStates}(c-d). The 2D asymmetric mode is stable to this 3D disturbance, whereas the 2D symmetric is unstable, although with only a very small growth rate. 
Additionally, for the 2D symmetric base flow, we obtain an unstable 2D mode. This mode is asymmetric and indicates that the 2D symmetric state transitions to the 2D asymmetric state under a small asymmetric perturbation.
This indicates that in 3D simulations the 2D asymmetric state is stable.
Next, we will investigate the possibility of additional steady states in 3D simulations.

\begin{table}
\centering
\begin{tabular}{l|c|c|c}
              2D Base Flow & Growth rate, $\mathrm{Re(\omega)}$ & Frequency, $\mathrm{Im(\omega)}$ & Dimensionality of eigenvector  \\ \hline
2D Asymmetric  & $-3.0243 \times 10^{-3}$ & 0 & 3D  \\
2D Symmetric & $1.2125 \times 10^{-4}$ & 0 & 3D \\
2D Symmetric & $2.56 \times 10^{-3}$ & 0 & 2D
\end{tabular}
\caption{\label{tab:table2} Growth rate and frequency for secondary linear stability analysis of 2D symmetric and 2D asymmetric steady states with $\Pi_s = 0.9$, $\Pi_h = 1500$, and $L_z = 1.6$.}
\end{table}

\subsection{Three-dimensional direct numerical simulations}

We conduct 3D DNS of the complete Navier-Stokes equations with varied initial disturbances to delve deeper into the behavior of the 3D symmetric eigenmode identified through LSA. Conventionally, it is anticipated that the 3D eigenmode would progress towards a corresponding 3D state, akin to the evolution observed in the 2D instabilities. This progression introduces an additional spatial dimension to the final flow state, thereby increasing its complexity. However, our findings suggest that this is not the case here.


We perform DNS for the case with $\Pi_s = 0.9$, $\Pi_h = 1500$, and $L_z = 1.6$, the optimal $L_z$ for the 3D symmetric eigenmode. 
\textcolor{black}{As previously explained, this set of parameter values was chosen because it is relatively close to the critical value for the 3D instability, and thus exhibits an orderly flow transition that is convenient to analyze in depth. For context, at a fixed $\Pi_h = 1500$, the critical $\Pi_s$ for the 3D symmetric instability is approximately 0.871, and a similar transition scenario is seen up until a $\Pi_s$ of approximately 1.3 to 1.4. The Reynolds number, defined by the maximum $u$ velocity of the final state and the height of the valley, is approximately 3.6, whereas at $\Pi_s = 1.4$, the Reynolds number is approximately 15. }

\begin{figure*}
\centering
\includegraphics[width=0.55\textwidth]{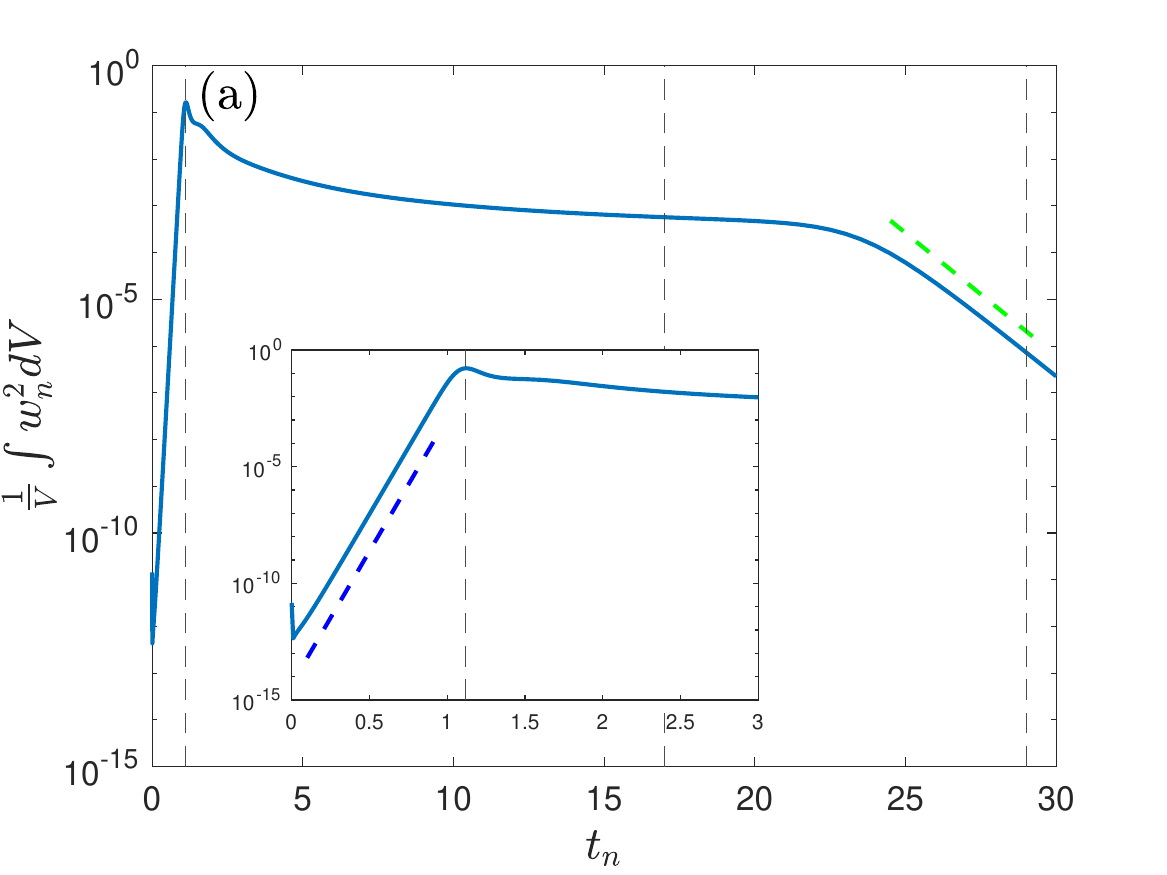} 
\includegraphics[width=0.3\textwidth]{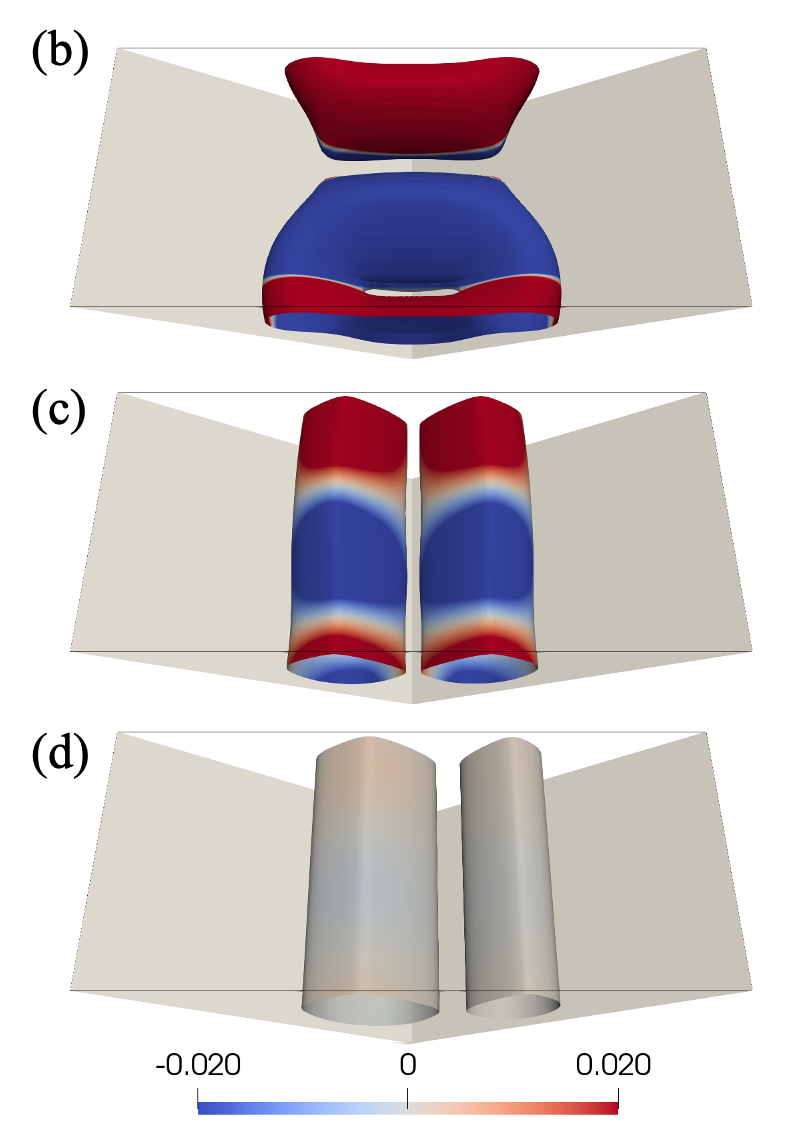}
\caption{\label{fig:wSquared} (a) Volume-averaged $w^2$ velocity over time for $\Pi_s = 0.9, \Pi_h = 1500$, and $L_z = 1.6$, along with growth/decay rates predicted by LSA at two different points in the evolution shown with colored dashed lines. The inset shows the evolution during early times. Vertical dashed lines represent the times at which the flow field is depicted with the Q-criterion in (b) $t_n = 1.12$, (c) $t_n = 17$, and (d) $t_n = 29$, taken as 4\% of the maximum, and are colored with $w$ velocity. }
\end{figure*}

The simulation is started with the pure conduction base state plus an initial perturbation defined by a small amplitude random noise such that the maximum amplitude of the disturbance is approximately $10^{-5}$ times the maximum amplitude of the final steady state. 
The time evolution of the volume averaged $w^2$ is shown in Figure \ref{fig:wSquared}(a) indicating the degree of three-dimensionality of the flow over time. Interestingly, following the nonlinear saturation of the initial exponential growth, the flow does not evolve into an expected  3D state. Instead, the $w$-velocity begins to decay, and the flow ultimately settles into a 2D steady state. 
Visualization of the flow field at three different points in the evolution are shown in Figs. \ref{fig:wSquared}(b-d) with the Q-criterion. At $t_n = 1.12$, the flow is at the maximum three-dimensionality and forms a strong vortex ring circulation structure, as depicted in Figure \ref{fig:wSquared}(b). After the nonlinear saturation of the initial exponential growth of the instability, there is a brief period of fast decay of $w$, followed by a long period of slow decay, from approximately $t_n = 5$ to $t_n = 22$. During this period, the flow field comes to resemble the 2D symmetric state with a small and diminishing magnitude of $w$ velocity, as shown in Figure \ref{fig:wSquared}(c). Finally, at later times, for $t_n > 24$, the decay of the $w$ velocity suddenly increases to the point of $w$ approaching zero. This is associated with the 2D symmetric state transitioning to the 2D asymmetric state, as shown in Figure \ref{fig:wSquared}(d).
This evolution represents an unstable 3D instability that spontaneously self-organizes to a less complex 2D steady state. \textcolor{black}{In order to determine if the transient 3D state is a possible equilibrium state, we run additional simulations restarted from the maximum 3D amplitude with smaller timestep and lower $\Pi_s$, but in each case we observe a decay of the $w$ velocity, as described here.}

To unravel the reasons underlying this behavior, we separate the evolution into four distinct phases. The first phase from the start of the simulations to $t_n = 1.12$ represents the phase of linear growth. This phase is shown in more detail in the inset of Figure \ref{fig:wSquared}(a), in which the growth rate of the 3D eigenmode is plotted \textcolor{black}{by a blue dashed line} alongside the initial exponential growth of the simulation, and these growth rates match well. Therefore, this phase follows the expected linear stability behavior, before any nonlinear effects become significant. During this phase, the velocity profile evolves from the shape of the 3D symmetric eigenmode, shown in Figure \ref{fig:eigenvector}(a), to the state with the maximum $w$ velocity shown in Figure \ref{fig:wSquared}(b). 

The second phase represents the nonlinear saturation of the instability, and is manifested by the fast decay of the $w$ velocity immediately after the peak at $t_n = 1.12$ until approximately $t_n = 5$. In traditional transition to turbulence theory, after the nonlinear saturation of the instability, the flow field evolves into a new steady or oscillating state upon which secondary instabilities can arise \citep{schmid2002stability}. 
However, we see an initial fast decay of the $w$ velocity which significantly changes the structure of the flow field. 
We investigate the cause of this decay through budget analysis of the disturbance kinetic energy (DKE) equation. We define the perturbation velocity $\mathbf{u^{\prime}}$ and the perturbation buoyancy $b^{\prime}$ as the difference between the flow field and the motionless base state defined in Eq. \ref{eq:base_flow}. 
Because the base flow is motionless, the disturbance velocity is equal to the instantaneous velocity at any point in time, so we drop the prime from this term.
Following the formulation of \citet{joseph1965stability}, the DKE equation can be written as
\begin{equation}
    \frac{d E_K}{d t} = - \int_{V} \left( \mathbf{u} \cdot \mathbf{D} \cdot \mathbf{u} + \mathbf{g} \cdot \mathbf{u} b^{\prime} + \nu \nabla\mathbf{u} : \nabla\mathbf{u}  \right) dV, 
\end{equation}
where $E_K = 0.5 \int_V \left( \mathbf{u} \cdot \mathbf{u} \right) dV$, $\mathbf{D}$ represents the strain rate tensor of the base flow, \textcolor{black}{and $\mathbf{g}$ is the effective gravity vector $\mathbf{g} = [0, 1, 0]$ representing the direction of gravity.} Because our base flow has zero velocity and gravity acts only in the $y$ direction, we can simplify the above equation to the following:
\begin{equation} \label{eq:ke}
    \frac{d E_K}{d t} = \underbrace{ - \int_{V} v b^{\prime} dV}_\text{Buoyant production} \underbrace{- \int_V \nu \nabla\mathbf{u} : \nabla\mathbf{u} dV}_\text{Viscous dissipation}.
\end{equation}
The budget analysis of the DKE equation simplifies to the interplay of two terms: the first signifies buoyant production, leading to an increase in DKE, while the second denotes DKE dissipation due to viscous effects.

\begin{figure*}
\centering
\includegraphics[width=0.45\textwidth]{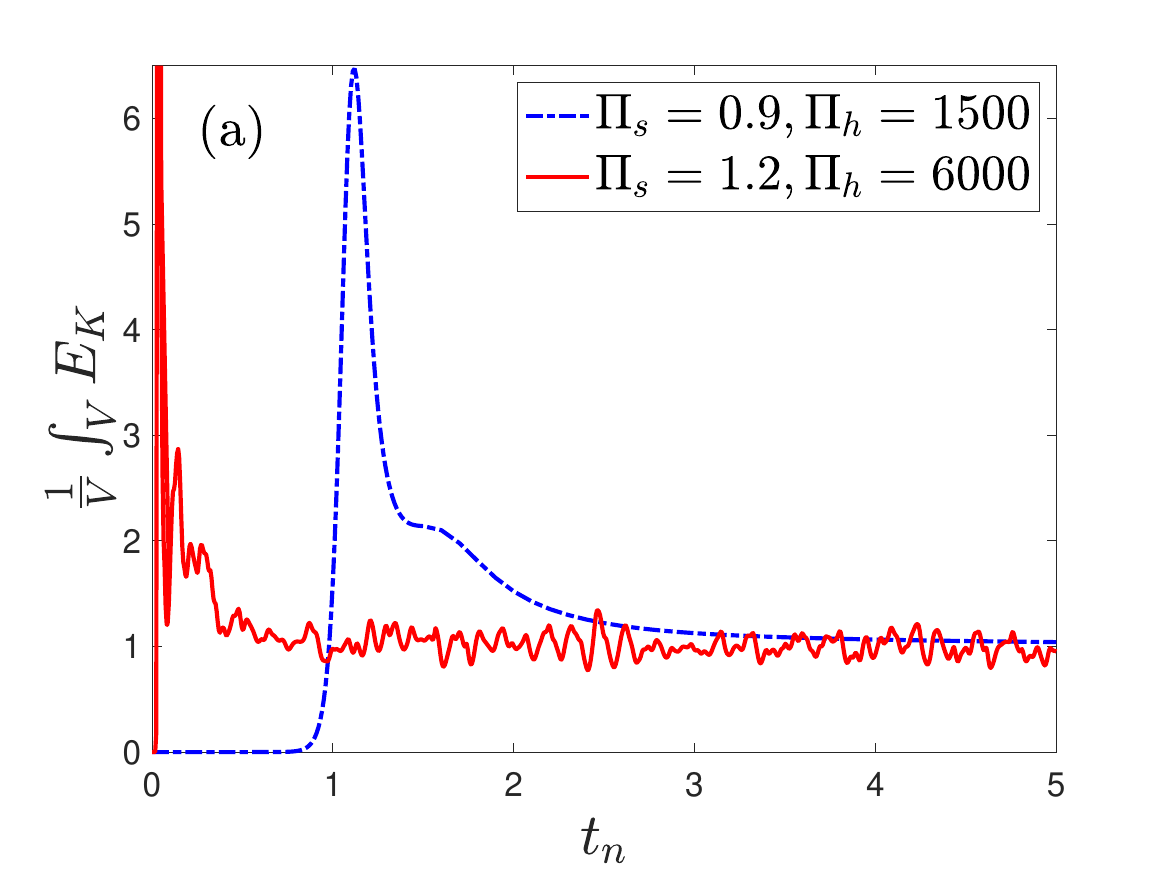}  
\includegraphics[width=0.45\textwidth]{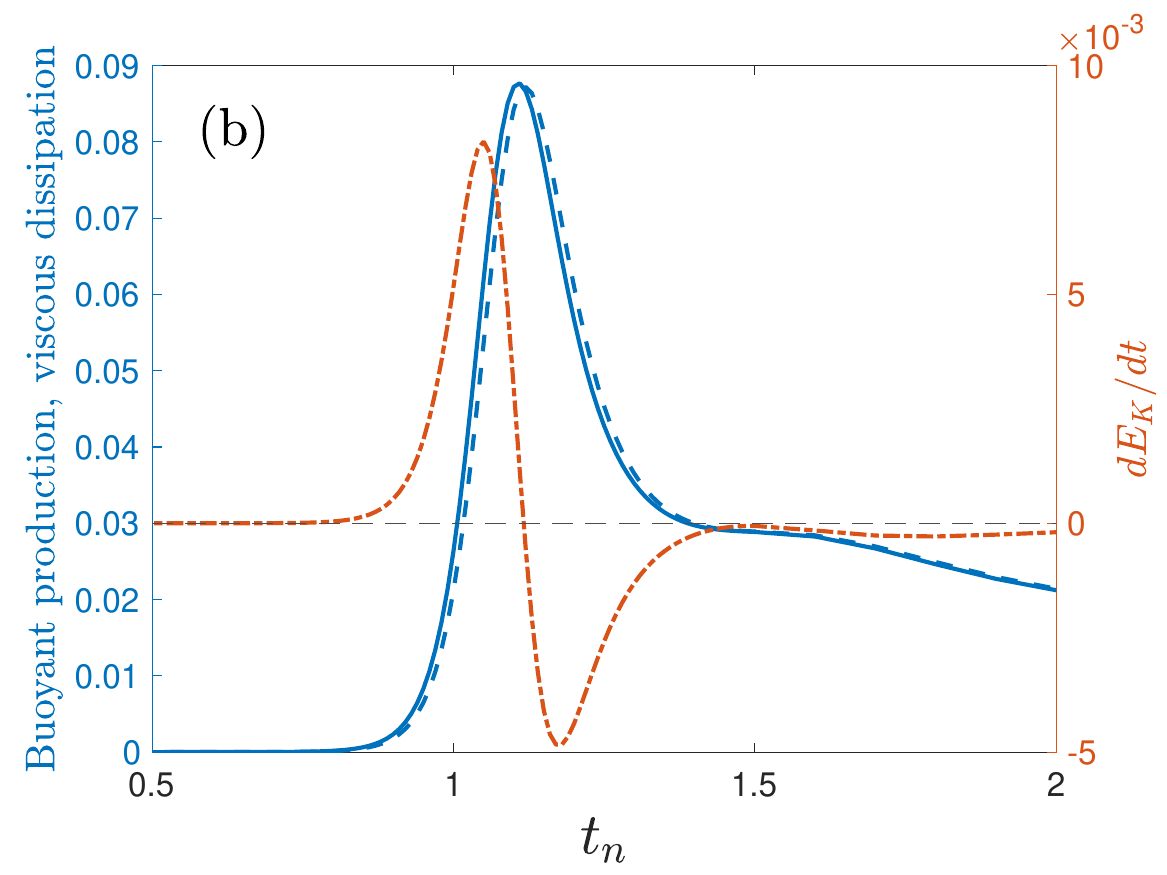}
\includegraphics[width=0.45\textwidth]{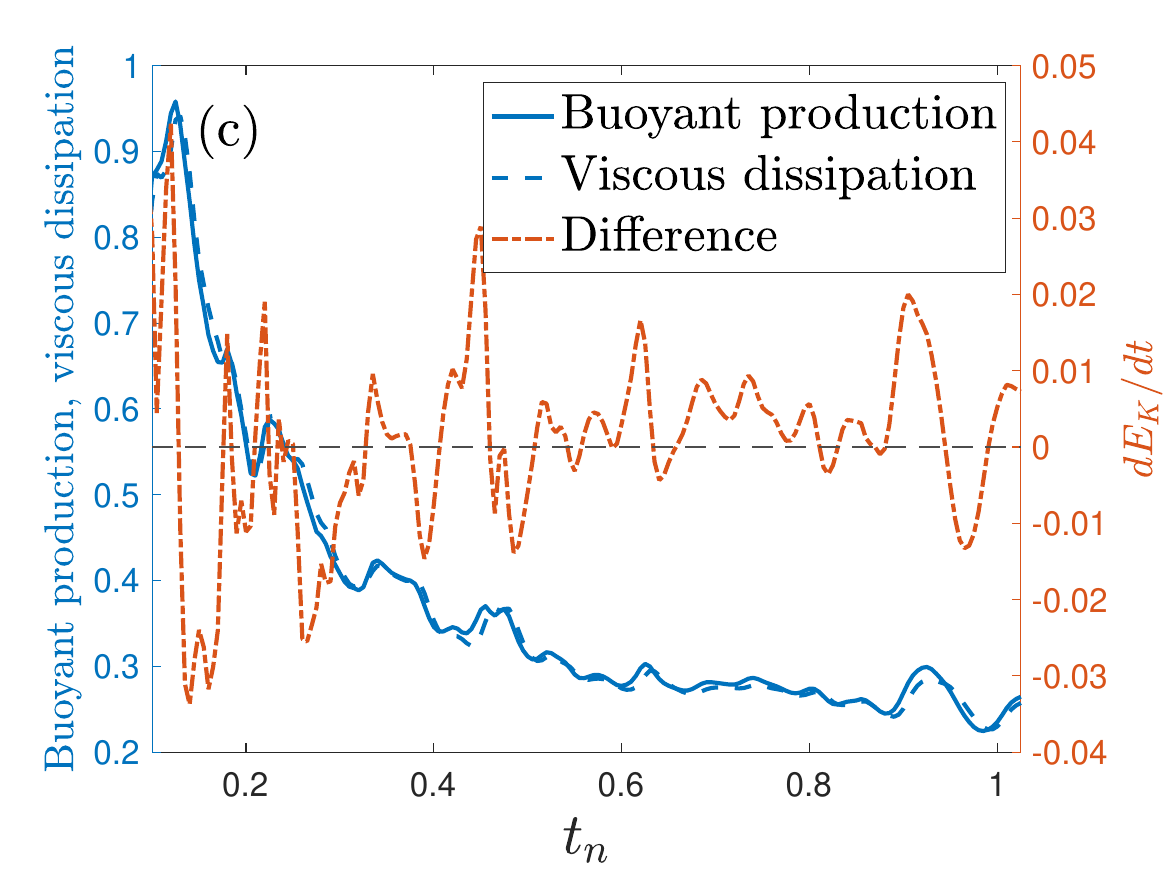}
\caption{\label{fig:KEterms} (a) Evolution of total DKE over time for two cases: $\Pi_s = 0.9, \Pi_h = 1500$ and $\Pi_s = 1.2, \Pi_h = 6000$. $L_z = 2$ in both cases. (b) Evolution of (absolute value of) buoyant production and viscous dissipation terms (left axis) and \textcolor{black}{the sum of buoyant production and viscous dissipation, or $dE_K/dt$} (right axis) for $\Pi_s = 0.9, \Pi_h = 1500$ case. (c) shows the same as (b) for the $\Pi_s = 1.2, \Pi_h = 6000$ case. \textcolor{black}{The DKE values in (a) are normalized by final statistically steady state values.}}
\end{figure*}

Figure \ref{fig:KEterms} shows the DKE budget terms for two different cases. The first case is with $\Pi_s = 0.9$ and $\Pi_h = 1500$, which is the same case pictured in Figure \ref{fig:wSquared} and exhibits the 3D instability self-organizing to the 2D steady state. In order to contrast this with a more unstable case, we compare it to a simulation with $\Pi_s = 1.2$ and $\Pi_h = 6000$, which remains 3D and unsteady for the entire evolution. This can be seen in Figure \ref{fig:KEterms}(a) which shows the evolution of DKE for both cases. \textcolor{black}{The DKE of each case in Figure \ref{fig:KEterms}(a) is normalized by its statistically steady state value in order to see the trend of both curves. In absolute value, the final DKE of the 3D unsteady state is approximately two orders of magnitude larger than the transient 3D state.}

Figure \ref{fig:KEterms}(b) and \ref{fig:KEterms}(c) show the DKE budget equation terms versus time for the early evolution for the $\Pi_s = 0.9$ and $\Pi_s = 1.2$ cases, respectively. Focusing first on the $\Pi_s = 0.9$ case, we see that the absolute value of the buoyant production and viscous dissipation terms follow a very similar trend. However, we note that there is a small difference between the two curves that is critical for explaining the observed self-organization. 
\textcolor{black}{The sum of buoyant production and viscous dissipation, equivalent to $dE_K / dt$, is plotted on the right vertical axis. This is also equivalent to the difference between the two curves because the absolute magnitudes of each term are plotted in Figure \ref{fig:KEterms}. }
When this sum is positive, it indicates that the buoyant production exceeds the viscous dissipation, and $dE_K / dt$ increases. When negative, the viscous dissipation exceeds the buoyant production and $dE_K / dt$ decreases. From Figure \ref{fig:KEterms}(b), immediately before the peak in DKE at $t_n = 1.12$, the buoyancy term dominates and causes the DKE to increase. However, at the peak of the KE, the viscous dissipation matches and then exceeds the buoyant production term, causing the DKE to decrease with time. 
Thus, the driving mechanism of the initial exponential growth of the instability is the buoyant production, and the nonlinear saturation of this instability and subsequent decay is due to the increase of the viscous dissipation as the velocity field continues to evolve. We also note that the viscous dissipation aligns well with the diffusion timescale, causing the velocity to peak at a dimensionless time close to 1. While this is not observed for all cases, it provides further evidence that the nonlinear saturation of the instability depends upon the timescale of the diffusion to ``kick in" and cause the initial decay of the $w$ velocity.

The behavior of the $\Pi_s = 0.9$ case can be contrasted to the fully 3D, unsteady case with $\Pi_s = 1.2$, shown in Figure \ref{fig:KEterms}(c). Whereas for the previous case, there is one change in the dominant term of the budget equation, which represents the end of the growth of the instability and the initial dissipation of the DKE, this is not the case for the unsteady state, in which there is a constant competition between the two budget terms, neither of which is seen to dominate the other, leading to a 3D, unsteady flow state.

\begin{figure*}
\centering
\includegraphics[width=0.45\textwidth]{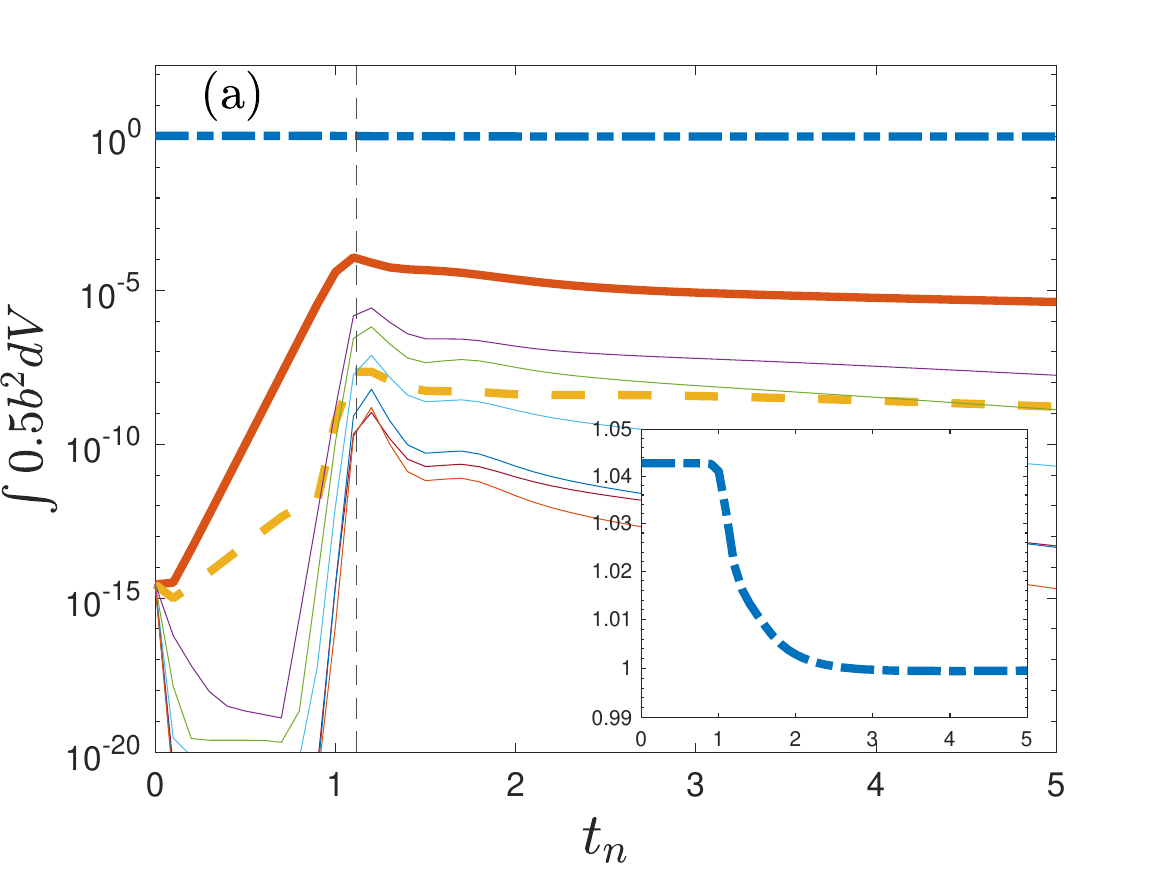}
\includegraphics[width=0.45\textwidth]{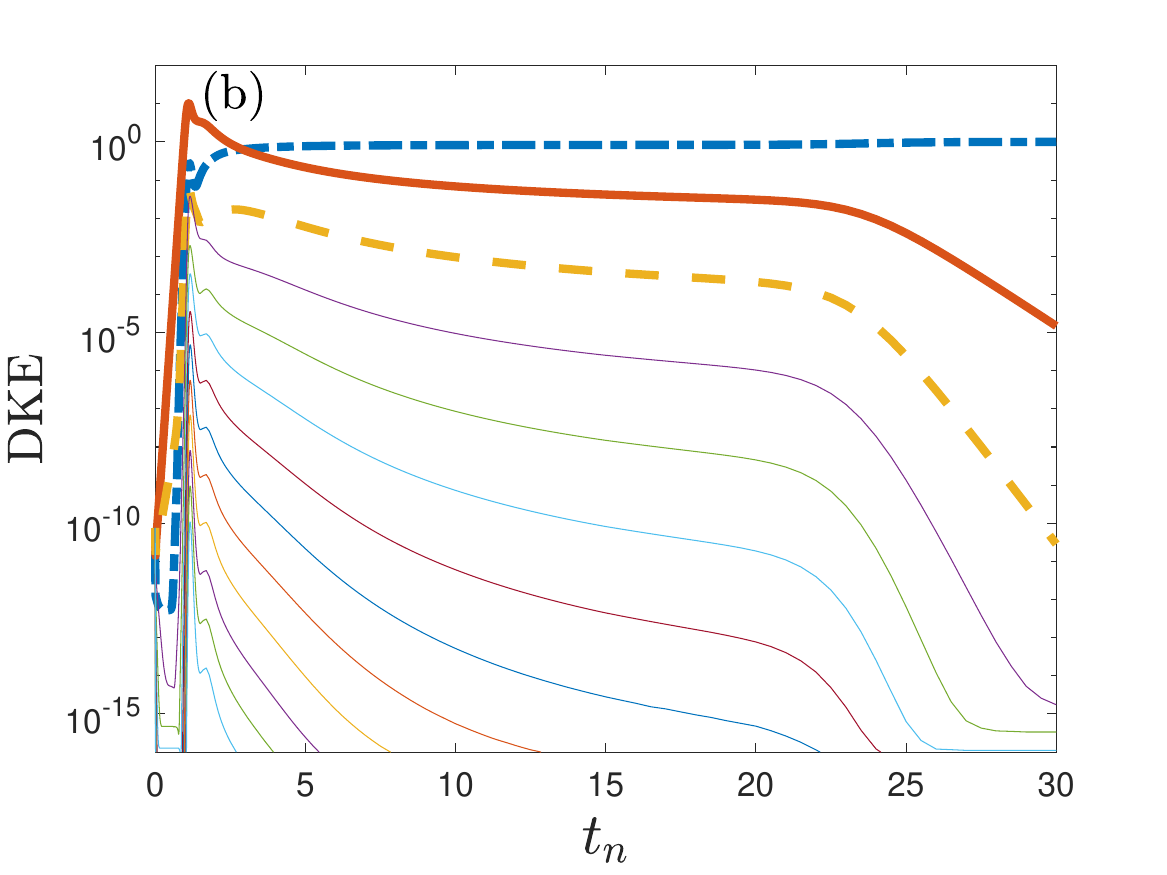}  
\includegraphics[width=0.45\textwidth]{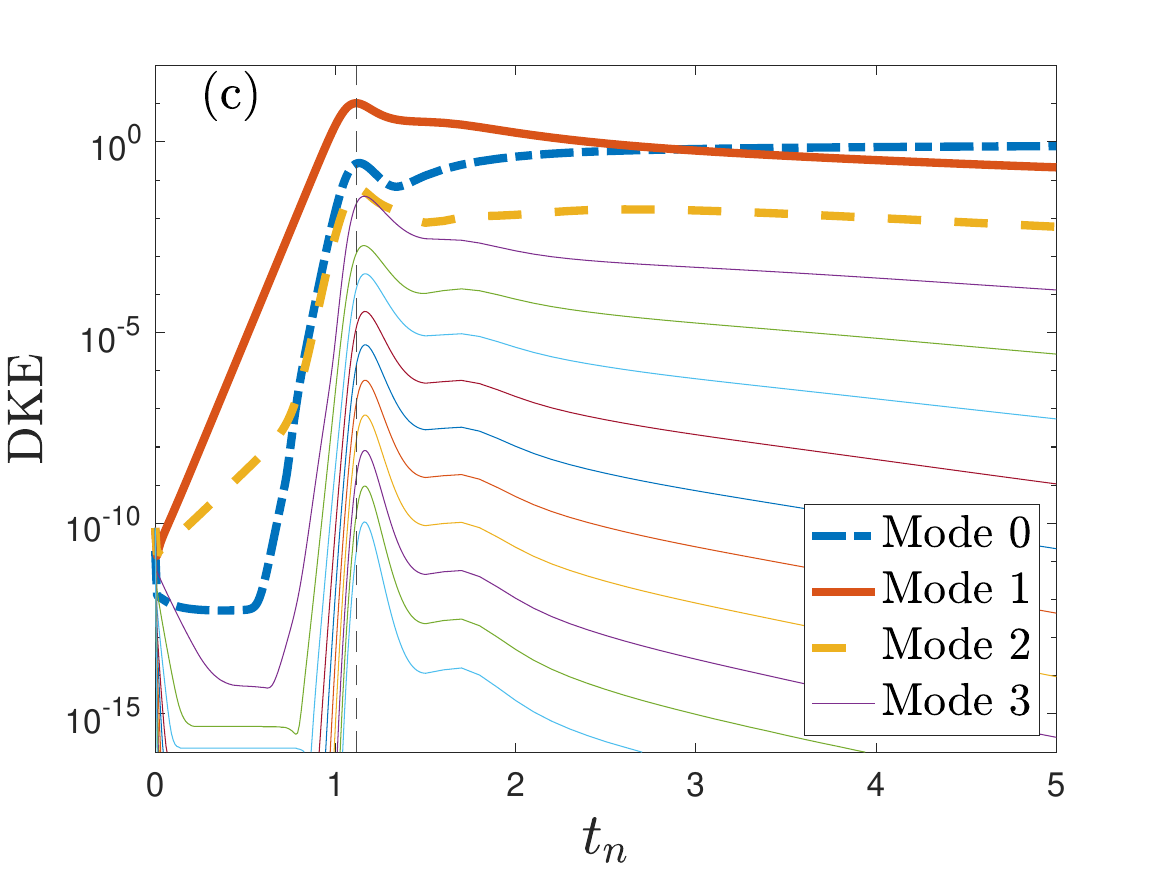}
\includegraphics[width=0.45\textwidth]{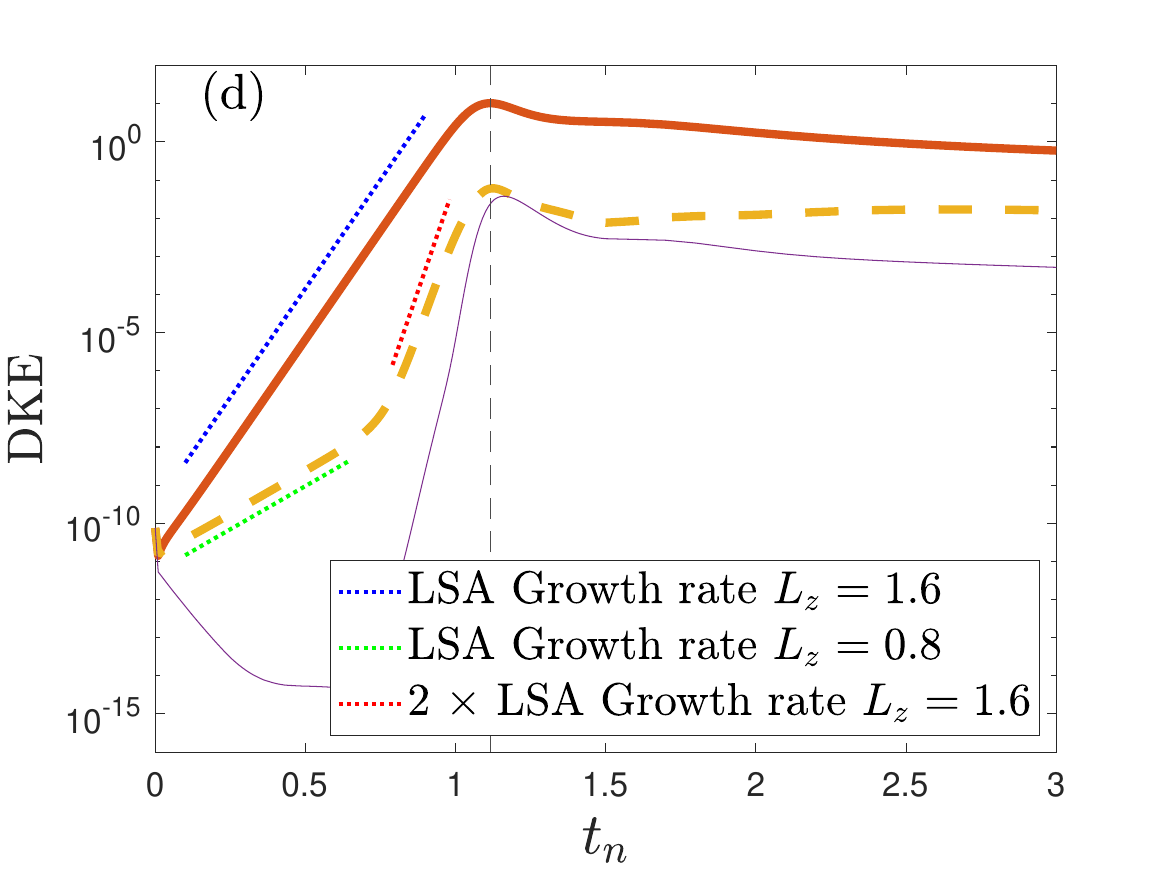}
\caption{(a) Evolution of the total buoyancy ($\int 0.5 b^2 dV$) for each Fourier mode versus time, with inset showing evolution of Mode 0 for early times. (b-d) Evolution of disturbance kinetic energy (DKE) for each Fourier component over time from the case with $\Pi_s = 0.9$ and $L_z = 1.6$ with initial random noise disturbance. The thicker lines represent the zero mode (or the mean flow), the first mode, and the second mode, and additional lines represent higher spanwise wavenumber components. (b) shows DKE of each mode for the entire evolution, (c) shows the evolution zoomed into earlier times, and (d) shows the comparison between the initial growth of the DKE of the first, second, and third modes with the growth rate predicted by LSA for the 3D symmetric mode at $L_z = 1.6$ and $L_z = 0.8$. The DKE/total buoyancy of each mode is normalized by the DKE or total buoyancy of the final 2D steady state relative to the pure-conduction state. }\label{fig:FourierModes}
\end{figure*}

To gain deeper insights into the current self-organization behavior, we extract the Fourier components of the flow field along the homogeneous $z$-direction. Figure \ref{fig:FourierModes}(a) illustrates the temporal evolution of total buoyancy for each Fourier mode during the initial phase, demonstrating the transfer of thermal energy between modes over time. 
The first three modes are highlighted by thicker lines in Figure \ref{fig:FourierModes}, including the mode corresponding to the 2D spanwise-averaged mode (Mode 0), the mode corresponding to the eigenvector with $L_z = 1.6$ (Mode 1), as well as the mode corresponding to the wavelength $L_z = 0.8$ (Mode 2). 
Starting from an one-dimensional buoyancy profile, we  observe energy transfer from Mode 0, equivalent to the 2D spanwise averaged component, to the dominant eigenmode (Mode 1) at early times. This is mostly a linear mechanism, as evidenced by the agreement of the growth rate of Mode 1 with values from LSA. During this initial phase of linear growth of the 3D instability, the energy of Mode 1 and Mode 2 increase, whereas at later times $t_n>0.6$ and  before the eigenmode reaches its maximal amplitude at $t_n=1$, higher wavenumber modes begin to increase exponentially with a similar growth rate as the eigenmode due to nonlinear interactions with the mean mode.  The inset of Figure \ref{fig:FourierModes}(a) shows the decay of Mode 0 of the temperature field which occurs at around $t_n = 1$, when all higher modes also begin to decay, closely correlating with the nonlinear saturation of the instability and the growth of mode 0 of the velocity field. This suggests that thermal energy is being converted to mean kinetic energy and also dissipated by viscosity at small scales generated via the higher-order modes. It is worth notice that even at their maximal amplitude, all modes with higher wave numbers than the eigenmode are still orders of magnitude weaker than the mean mode, indicative of a very small short-wavelength correction to the 2D mean flow field. 

In Figure \ref{fig:FourierModes}(b-d) we plot the total disturbance kinetic energy (DKE) relative to the pure-conduction state for each Fourier mode, with the DKE of each mode normalized by the DKE of the final 2D steady state. 
Because we begin our simulation from an initial condition of small amplitude random noise added to the zero base flow, the DKE magnitude of all modes begin at approximately the same value. During the initial period of the simulation, the period of linear growth, Mode 1 undergoes fast exponential growth, while Mode 2, depicted with the dashed line undergoes comparatively slower exponential growth. All other modes initially decay. Because this phase represents the period of linear growth, we can explain this behavior through the results of our LSA.

Figure \ref{fig:FourierModes}(d) shows the initial growth of the DKE of Mode 1 and Mode 2 along with the growth rates from LSA for the 3D symmetric eigenvector at $L_z = 1.6$ and $L_z = 0.8$. We see that the growth rate of the 3D symmetric eigenvector at $L_z = 1.6$ matches closely with the growth rate of the DKE of the first mode, and likewise the growth rate of the 3D symmetric eigenvector at $L_z = 0.8$ matches closely with the growth rate of the DKE of the second mode. From our LSA, we know that we obtain the unstable 3D symmetric eigenmode at both $L_z = 1.6$ and $L_z = 0.8$, and this explains the initial growth we see in the first and second Fourier modes. 
For higher spanwise wavenumbers, the base state is linearly stable to any 3D disturbance, and thus we see initial decay in higher wavenumber modes before dimensionless time of approximately 0.5.

As the period of linear growth nears its end and nonlinear saturation of the instability begins, we see the other Fourier modes begin to exhibit exponential growth along with an increase in the rate of growth of Mode 2, which continues until after the peak in the energy of Mode 1. The amplification of the higher spanwise wavenumber modes before the peak of DKE in Mode 1 reflects the growing nonlinearity in the simulations, and begins at approximately $t_n = 0.7$. This can be seen from the growth rate of mode 2 in Figure \ref{fig:FourierModes}(d); after about $t_n=0.7$, the growth rate of Mode 2 increases to approximately twice the growth rate of the fundamental mode. This increase of growth rate is consistent as well with the higher order modes in Figure \ref{fig:FourierModes}(c). Additionally, we note that the peak in DKE of Mode 1, marked by a vertical dashed line in Figure \ref{fig:FourierModes}(c) occurs shortly before the peak of the higher wavenumber modes. 

After the peak in the DKE, the energy of all the Fourier modes begins to decay besides Mode 0, or the mode representing the spanwise-averaged field. This indicates that the energy of all the higher wavenumber modes is being transferred to sustain the 2D flow, and that by $t_n = 3$, the DKE of Mode 0 exceeds Mode 1, meaning the energy of the 2D flow exceeds that of the initial dominant eigenvector. We also note how the decay rates of the 3D modes differ, with the higher wavenumber modes decaying faster, and how Mode 1 remains the dominant 3D mode in the flow field over the entire evolution. 

\begin{figure*}
\centering
\includegraphics[width=0.98\textwidth]{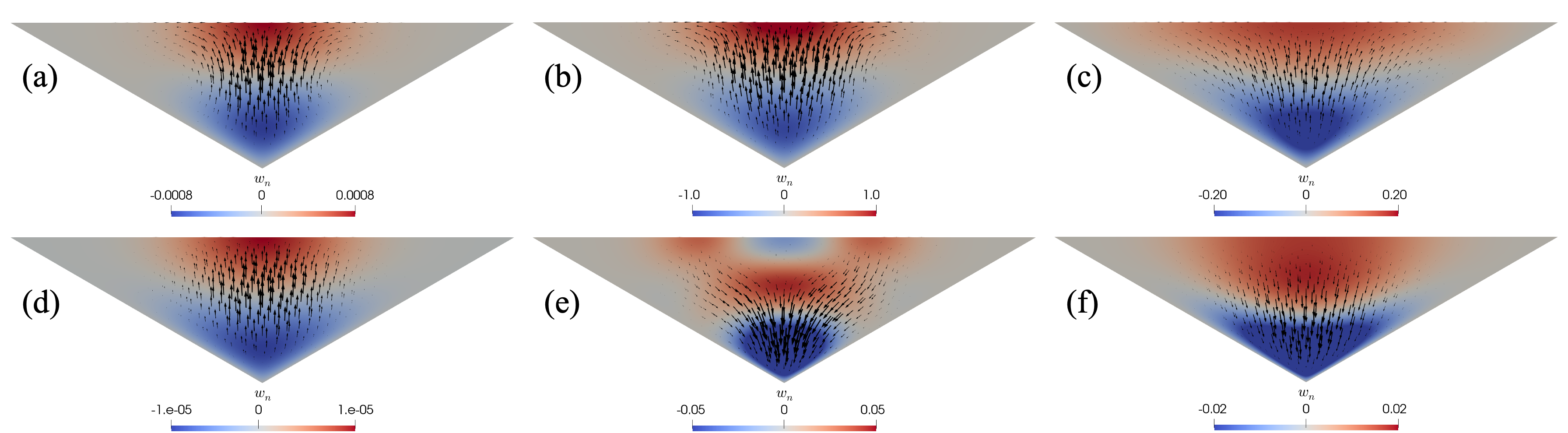}  
\caption{\label{fig:Mode1_Mode2} Visualization of the Fourier coefficients corresponding to Mode 1 at dimensionless times of (a) 0.5, (b) 1.12, and (c) 3, and Mode 2 at (d) 0.5, (e) 1.12, and (f) 3. The 2D flow field is colored by the $w$ velocity and velocity vectors depict the 2D $u$ and $v$ components.  } 
\end{figure*}

Figure \ref{fig:Mode1_Mode2} shows the 2D visualization of the Fourier coefficients corresponding to the two strongest 3D modes, Mode 1 and Mode 2, at three different times in the evolution. The Fourier coefficients for Mode 1, the dominant 3D mode in our simulation, are depicted in Figure \ref{fig:Mode1_Mode2}(a-c), with (a) depicting the flow at $t_n = 0.5$, still in the linear growth regime, (b) representing the saturation of the linear instability at $t_n = 1.12$, and (c) representing the flow after the initial decay at $t_n = 3$. The structure of Mode 1 does not change significantly during the evolution, with a strong upward flow of the 2D components along with a strong circulation in $y-z$ plane. However, at later times, after the energy of the mode decays, both the 2D and 3D components of velocity appear to diffuse throughout the width of the valley in the $x-y$ plane. 
The Fourier coefficients of Mode 2 are depicted for the same times in Figure \ref{fig:Mode1_Mode2}(d-f). The field during the linear growth phase matches closely with that of Mode 1, which makes sense as each represents the linear growth of the 3D symmetric eigenmode although with different wavelengths. However, unlike Mode 1, the structure of the Fourier coefficients of Mode 2 change significantly near the nonlinear saturation and after, as can be seen in Figure \ref{fig:Mode1_Mode2}(e) and (f), with the 2D flow eventually flipping directions from upward to downward after the beginning of the decay of the 3D structures. These changes can be explained by the nonlinear interactions with Mode 1 which cause the abrupt increase in the growth rate of higher spanwise wavenumber modes seen at about $t_n = 0.7$. Ultimately, similar to Mode 1, the flow field appears to diffuse throughout the valley during the decay of the DKE. 

\begin{figure*}
\centering
\includegraphics[width=0.98\textwidth]{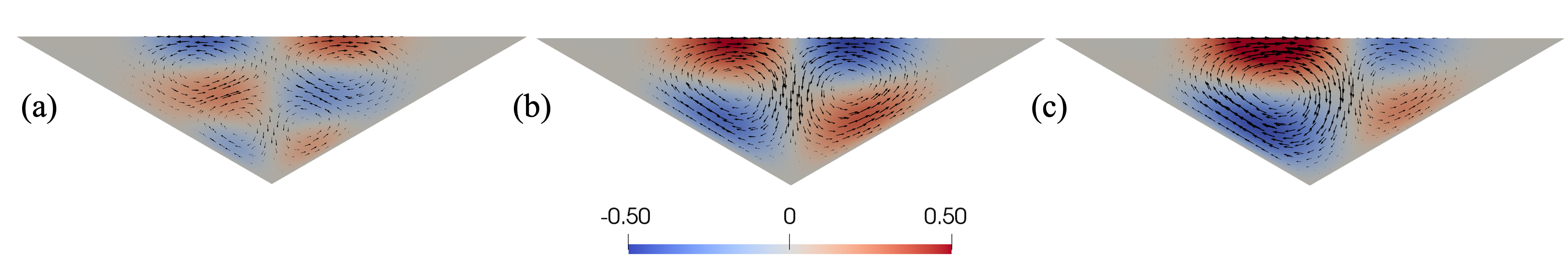}  
\caption{\label{fig:Mode0} Visualization of the 2D averaged flow field at dimensionless times of (a) 1.12, (b) 17, and (c) 29, corresponding the vertical dashed lines and 3D visualizations in Figure \ref{fig:wSquared}. Colored by $u$ velocity and arrows show velocity vectors.  } 
\end{figure*}

Figure \ref{fig:Mode0} shows the 2D spanwise-averaged flow field, equivalent to Mode 0, for three different times in the flow evolution. For $t_n = 1.12$, shown in Figure \ref{fig:Mode0}(a), the three-dimensionality of the flow field is at a maximum, and the spanwise-averaged field does not resemble any known 2D state in the valley. However, after the nonlinear saturation of the instability, and the decay of the 3D Fourier modes, by $t_n = 17$, the spanwise-averaged flow field shown in Figure \ref{fig:Mode0}(b) matches closely with the 2D symmetric steady state shown in Figure \ref{fig:2dSteadyStates}(a). Likewise, near the end of the evolution, the spanwise-averaged flow field shown in Figure \ref{fig:Mode0}(c) matches closely with the asymmetric steady state shown in Figure \ref{fig:2dSteadyStates}(b). This confirms that after the initial growth of the 3D instability, the flow field begins converging to the 2D symmetric flow state, and because the 2D symmetric flow state is unstable to the 2D asymmetric flow state, the flow field ultimately transitions to the 2D asymmetric flow state. 

\begin{figure}
\centering
\includegraphics[width=0.7\textwidth]{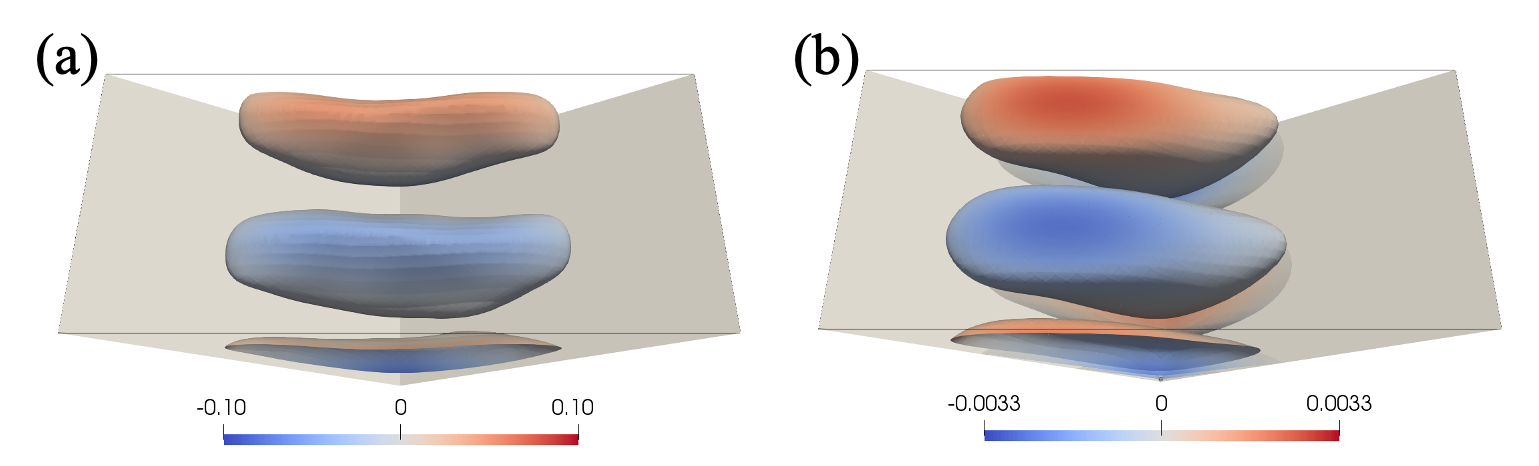}
\caption{\label{fig:SecStabAnalysis} Visualizations of the disturbance field of the simulation shown in Figure \ref{fig:wSquared} relative to  (a) the 2D symmetric base flow at $t_n = 17$ and (b) the 2D asymmetric base flow at $t_n = 29$. (a) is shown by contours of Q-criterion and (b) is contours of $x$-vorticity, both colored by $w$ velocity. \textcolor{black}{Note that $x$-vorticity is used in (b) and Figure \ref{fig:2dSteadyStates}(d) because the disturbance velocity field is too small at $t_n = 29$ to visualize the structures using Q-criterion.}}
\end{figure}

Associated with the transition to the 2D asymmetric state at approximately $t_n = 20$ is a large increase in the decay rate of the three-dimensionality, as shown by both $w^2$ in Figure \ref{fig:wSquared} and the 3D Fourier modes in Figure \ref{fig:FourierModes}. In order to explain this sudden increase in decay rate, we turn to the results of our secondary LSA described in Section \ref{sec:2d}. 
For 3D LSA with the 2D symmetric steady state as the base flow, we obtain one 2D eigenmode with a growth rate of $2.56 \times 10^{-3}$, representing the 2D asymmetric instability, and one 3D eigenmode with a growth rate of $1.21 \times 10^{-4}$, which is much smaller than the 2D growth rate. The 3D eigenvector was described earlier and is depicted in Figure \ref{fig:2dSteadyStates}(c). The positive growth rate of the 3D mode does not match with the slow decay of the $w$ velocity we observe in Figure \ref{fig:wSquared}(a) from $t_n$ of 10 to approximately 20, when the flow state is converging to the 2D symmetric flow state. To explain this, we visualize the disturbance flow field relative to the 2D symmetric flow state at $t_n = 17$ in our simulation, and this is shown in Figure \ref{fig:SecStabAnalysis}(a). While this resembles the eigenvector, there is also a clear asymmetry in the disturbance flow field, and this  asymmetric deviation helps reduce the growth rate of the disturbance's evolution below the already small growth predicted by LSA for the optimal symmetric eigenmode, and contributes to the increase of the unstable 2D asymmetric mode. In other words, the small asymmetry in the disturbance field limits the energy transfer from the symmetric 3D state to the dominant 3D eigenmode, which prevents the positive growth rate of any 3D disturbance as predicted by LSA, and instead transfers the energy to the 2D flow. Therefore, this explains the slow decay we see during the middle portion of the simulations from $t_n$ between 3 and 20.

When we perform 3D LSA with the 2D asymmetric steady state as the base flow, we find a 3D eigenmode with a negative growth rate of $-3.02 \times 10^{-3}$, as described previously in Section \ref{sec:2d}. This decay rate is shown in Figure \ref{fig:wSquared}(a) by the dashed green line. Unlike for the middle portion of the simulation, the decay rate matches well with what we observe in our simulation.
Additionally, we show the 3D flow structure of our simulation at $t_n = 29$ in Figure \ref{fig:SecStabAnalysis}(b), and we find good agreement in structure with the eigenvector of the secondary instability shown in Figure \ref{fig:2dSteadyStates}(d). This explains the rapid increase in the decay rate of the $w$ velocity after $t_n$ of about 24 in Figure \ref{fig:wSquared}(a).

\subsection{Domain with a long span-wise extent and with initial random disturbance}
\begin{figure*}
\centering
\includegraphics[width=0.9\textwidth]{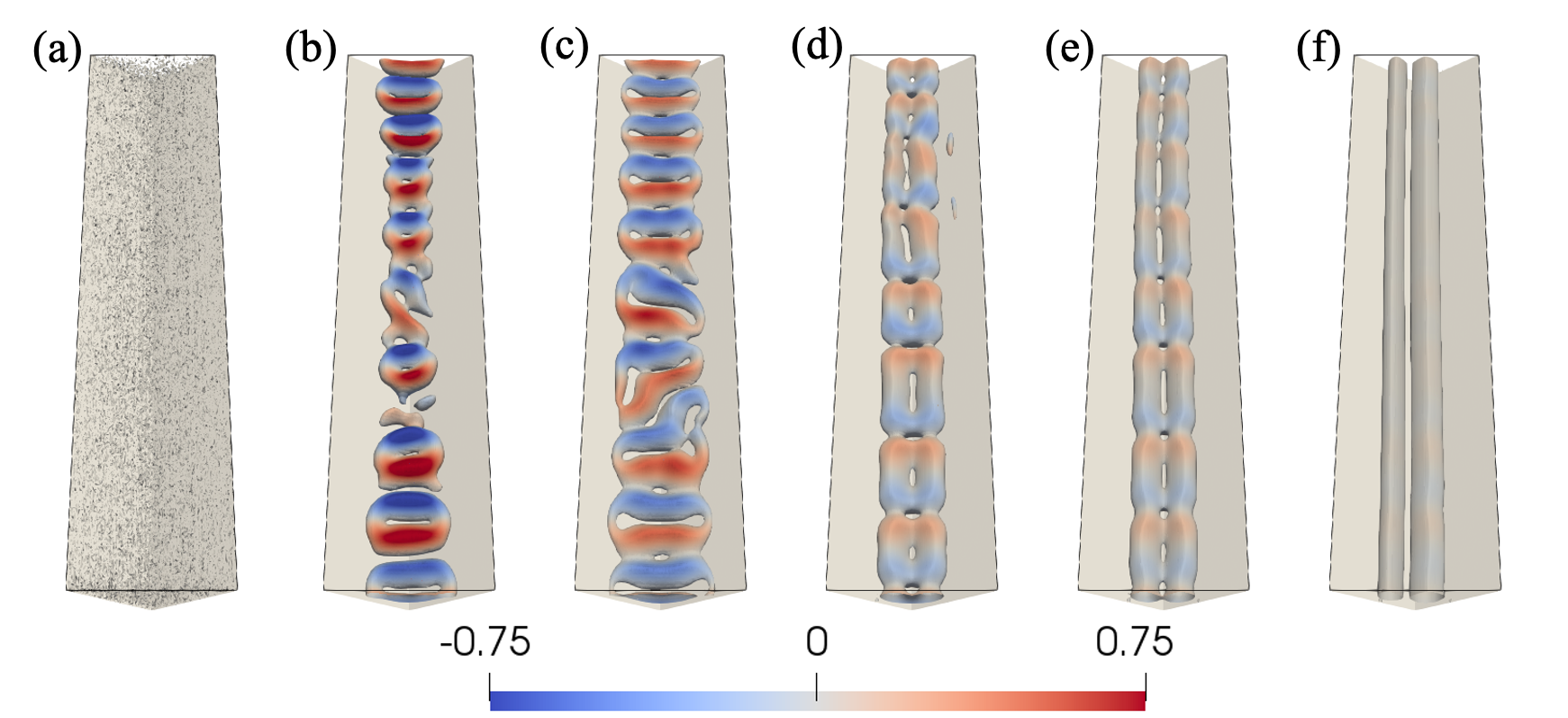}
\caption{\label{fig:Qcrit_16Lz} Visualization of the Q-criterion over time for DNS of the case $\Pi_s = 0.9$, $\Pi_h = 1500$, and $L_z = 16$. Initial conditions are given as a small amplitude random noise, depicted in (a). Subsequent non-dimensional times are (b) 1, (c) 1.5, (d) 2.5, (e) 4, and (f) 10. Contours of Q-criterion are taken as 4\% of the maximum, and are colored by the $w$ velocity. A movie of the full evolution is attached as supplementary Movie 2. }
\end{figure*}

We perform an additional simulation with $L_z = 16$ with an initial Gaussian random perturbation to ensure our results are not impacted by a limited extent in the $z$ direction. We observe the same spontaneous growth of a 3D instability before the 3D structures self-organize back to the final 2D steady state. Visualizations of the flow field at six different times in the evolution are shown in Figure \ref{fig:Qcrit_16Lz}. Movies of the full transition for both $L_z = 1.6$ and $L_z = 16$ cases are attached as supplementary Movie 1 and Movie 2. When comparing the flow structures pictured here to the flow structures observed in the $L_z = 1.6$ case, while they are similar (for example, compare Figure \ref{fig:Qcrit_16Lz}(b) and Figure \ref{fig:wSquared}(b)), we see that the structures in the $L_z = 16$ case are significantly more disordered, which we suspect is a consequence of the increased span in the homogeneous direction, which allows the structures to combine and interact which is restricted with an $L_z$ of only one wavelength of the instability. However, this more clearly shows the self-organization behavior as this relatively more disordered state transitions to the same ordered, 2D state as in the $L_z = 1.6$ case. The results of this simulation give us confidence that the spontaneous growth of 3D structures and subsequent decay of three-dimensionality is not influenced by the extent of the $z$ direction in our simulations. As previously discussed, these findings reinforce the conclusion that non-modal effects can be excluded in the flow system analysis.

\section{Conclusion}


We identified a self-organization behavior in a stably-stratified, V-shaped enclosure heated from below. The self-organization of the flow is characterized by the spontaneous onset of 3D motion from an initially quiescent flow state, which eventually evolves into a 2D steady state without any external intervention. This flow evolution persists consistently across random, 3D infinitesimal perturbations. The self-organizing dynamic arises due to interaction between various factors influencing the disturbance kinetic energy (DKE), as well as to the linearly stable nature of the resulting 2D steady state. Initially, buoyancy effects promote rapid growth of DKE in all three spatial dimensions. After a crucial phase, the 3D instability reaches nonlinear saturation and viscous dissipation of DKE takes over the buoyant production DKE, guiding the evolution of the DKE towards the eventual 2D steady state.

The present self-organization example is distinct from prior examples of self-organization in fluid systems. In our example, the flow field self-organizes from a 3D transient state into a 2D steady state in the absence of any external interference on the initial configuration.
From the quiescent, pure-conduction state of the stably stratified medium, LSA indicates an unstable 3D instability, suggesting a transition to a dynamically complex 3D flow state. However, Navier-Stokes simulations reveal a final 2D steady state, with the ``most natural" route from the initial state being a 3D path marked by self-organization. By most natural, we mean that given the base flow and any random 3D perturbation, the transition to the 2D steady state will always include the initial three-dimensionality and subsequent self-organization. While the 2D asymmetric steady state can also be reached starting from the 2D asymmetric eigenmode, this requires limiting the span in the z-direction to preclude the 3D instability from growing. All the dynamics observed in our case manifest spontaneously from an initial state with infinitesimal perturbations, and with no external forcing to cause this self-organizing behavior. This is in contrast to the examples of self-organization discussed in the Introduction as they rely on some significant initial disturbance to the flow (such as in decaying turbulence \citep{clercx1998spontaneous} or a suddenly stopped rotating cylinder \citep{kaiser2020stages}) or some external forcing to favor 2D flow such as rotation or a magnetic field \citep{zhao2014transition}. \textcolor{black}{While our 
choice of a valley geometry that is homogeneous in the spanwise direction may favor the formation of 2D rolls over 3D structures, this differs from any kind of external interference with the flow because the geometry does not change throughout the evolution, and LSA indicates a dominant 3D instability within the same valley geometry.} 

\textcolor{black}{Our case is seemingly similar to transient chaos in that it passes through a more complex transient state before settling into a final 2D attractor, but there are significant differences. Notably, our example cannot be classified as transient chaos because the flow remains entirely laminar throughout its evolution. 
Whereas transient chaos is caused by a non-attracting chaotic set, or a chaotic saddle, \cite{rempel2007origin}, our observed transient 3D state is better described as an unstable equilibrium state, similar to the 2D symmetric state, both of which eventually transition to the 2D asymmetric state, which is the only stable attractor in the phase space.
This distinction is reinforced by the fact that the entire transition pathway is well described by linear stability analysis, including the initial exponential growth of the transient 3D state and the final exponential decay to the 2D asymmetric state. Furthermore, our case cannot be defined as transient chaos because we do not observe sensitivity with respect to initial conditions; instead, we observe the same transition route through the 3D state for any arbitrary 3D disturbance.}

The insights gained from this example of self-organization have the potential to enhance our understanding of flow dynamics in geophysical and astrophysical systems. Furthermore, our identification of the mechanisms driving this self-organization could provide valuable guidance for developing control strategies to manipulate the flow state in convective systems.

\newpage
\bibliography{apssamp}

\end{document}